\documentclass[12pt]{article}
\usepackage{amsmath}
\usepackage{epsfig}
\def\fun#1#2{\lower3.6pt\vbox{\baselineskip0pt\lineskip.9pt
\ialign{$\mathsurround=0pt#1\hfil##\hfil$\crcr#2\crcr\sim\crcr}}}

\begin{document}
\title{ Equilibrium of nuclear matter in QCD sum rules
}
\author{E.G. Drukarev, M.G. Ryskin, V.A. Sadovnikova\\
{\em National Research Center "Kurchatov Institute"}\\
{\em B. P. Konstantinov Petersburg Nuclear Physics Institute}\\
{\em Gatchina, St. Petersburg 188300, Russia}}
\date{}
\maketitle

\begin{abstract}
We calculate the nucleon parameters in symmetric nuclear matter  employing the QCD sum rules
approach. We focus on the  average energy per nucleon and study the equilibrium states of the matter.
We treat the matter as a relativistic system of interacting nucleons. Assuming the dependence of the nucleon mass on the light quark mass $m_q$ to be more important than that of nucleon interactions we find that the contribution of the relativistic nucleons to the scalar quark condensate can be expressed as that caused by free nucleons at rest multiplied by the density dependent factor $F(\rho)$. We demonstrate that
there are no equilibrium states while we include only the condensates with dimension $d \leq 3$. There are equilibrium states if we include the lowest order radiative corrections and the condensates with $d \leq 4$. They manifest themselves for the nucleon sigma term $\sigma_N >60$ MeV. Including the condensates with
$ d \leq 6$ we  find equilibrium states of nuclear matter for $\sigma_N> 41$ MeV. In all cases the equilibrium states are due to influence of the relativistic motion of the nucleons composing the matter on the scalar quark condensate.
\end{abstract}

\section{Introduction}

The vacuum QCD sum rules enable to express the characteristics of the observed hadrons in terms of the
vacuum expectation values of the QCD operators, also known as condensates. The method was employed for
calculation of the meson and nucleon characteristics \cite{1}, \cite{2} (see also \cite{3}). Later the approach was extended for the case of finite baryon density. The list of references is presented in the review paper \cite{4}. The idea of QCD sum rules in nuclear matter is
to express the self energies of  {\em the probe proton} in terms of the in-medium QCD condensates.

We  gave a  detailed description of  the  approach in  \cite{4}. Here we just recall the main points. In the vacuum sum rules one starts with building of the function $\Pi^{(0)}(q)$ which describes the time-space propagation
 of the three quark system with the proton quantum numbers carrying the four-momentum $q$. It contains two scalar functions $\Pi^{i(0)}(q^2)$ ($i=q,I$), and  $\Pi^{(0)}(q)=\hat q\Pi^{q(0)}(q^2)+I\Pi^{I(0)}(q^2)$ with $\hat q= q_{\mu}\gamma^{\mu}$ while $I$ is the unit $4\times 4$ matrix.
Analysis of the dispersion relations for the functions $\Pi^{i(0)}(q^2)$ is the next step. They are considered at large and negative values of $q^2$. The asymptotic freedom of QCD enables to present the functions $\Pi^{i(0)}(q^2)$  as the  power series of $1/q^2$ and of the QCD coupling $\alpha_s$. The QCD condensates of are the coefficients of $1/q^2$  series, known as the Operator Product Expansion, OPE \cite{5}.
In the OPE the large distances contribution is included in terms of the vacuum expectation values of the local quark and gluon operators. The scalar quark condensate $\langle 0|\bar q(0)q(0)|0\rangle$ is the most important one.

The imaginary parts of the functions $\Pi^{i(0)}(q^2)$ are caused by  their singularities corresponding to real states with the quantum numbers of the proton. In the standard approach to description of the spectrum of the functions $\Pi^{i(0)}(q^2)$ the lowest lying pole (the proton) is written explicitly while the higher excited states are treated approximately. This is known as  the "pole+continuum" model \cite{1}-\cite{3}. The position of the nucleon pole $m$, its residue $\lambda^2$ and the effective continuum threshold $W^2$ are the unknowns of the QCD sum rules equations. Thus the vacuum QCD sum rules express the characteristics of free proton in terms of vacuum
QCD condensates.

In nuclear matter we introduce  the 4-vector $P=(m,{\bf 0})$ with $m$ being the vacuum value of the nucleon mass (we neglect the neutron-proton mass splitting). The function $\Pi(q,P)$ describes the time-space propagation in nuclear matter
 of the three quark system with the proton quantum numbers carrying the four-momentum $q$.There are three structures of the function $\Pi(q)$ proportional to the matrices $\hat q$, $\hat P$ and $I$, i.e. $\Pi(q,p)= \hat q\Pi^q(q,P)+\hat P\Pi^P(q,P)+I\Pi^I(q,P)$.  To separate the singularities connected with the probe nucleon from those connected with the matter itself we fix the value $s=(q+P)^2$ putting $s=4m^2$. Thus we consider $\Pi^i(q,P)=\Pi^i(q^2,s)$ with $i=q,P,I$ and analyze the dispersion relations in $q^2$ for the functions
$\Pi_i(q^2,s)$. They can be expanded in powers of $1/q^2$. The in-medium QCD condensates are the coefficients of the expansion. Imaginary part of the functions $\Pi^i(q^2,s)$ is treated in the "pole+continuum" model. The nucleon Dirac effective mass $m^*$ (or the scalar self energy $\Sigma_S=m^*-m$), the vector self energy  $\Sigma_V$,
as well as in-medium values of the nucleon residue $\lambda^2_{m}$ and of continuum threshold $W_m^2$ are the unknowns to be determined from
the QCD sum rules. Hence the finite density QCD sum rules express the characteristics of the nucleon in nuclear matter $m^*$ and $\Sigma_V$ in terms of the in-medium QCD condensates. The latter are the functions of the nuclear matter density $\rho$. Thus we expect to find the density dependence of
$m^*$ and $\Sigma_V$.

In the OPE and $\alpha_s$ expansions weakly correlated quarks are exchanged between our three quark system and the matter. The scalar and vector nucleon self energies can be viewed as due to exchange of mesons (systems of strongly correlated quarks) between the probe nucleon and the nucleons of the matter. Thus the exchange by systems of strongly correlated quarks between the probe nucleon and the nucleons of the matter is expressed in terms of exchanges by weakly correlated quarks with the same quantum numbers between the current and the matter.

We try to find if the finite density QCD sum rules approach provide the equilibrium of nuclear matter. To answer the question we calculate the average binding energy per nucleon ${\cal B}(\rho)$ in terms of $m^*(\rho)$ and $\Sigma_V(\rho)$. We look if the condition $d{\cal B}(\rho)/d\rho=0$ can be satisfied for a reasonable value of $\rho=\rho_{eq}$ with ${\cal B}(\rho_{eq})<0$.

The convergence of $1/q^2$ series is an important assumption. Thus we expect the terms containing the condensates of the lowest dimension to be the most important ones. The higher terms of the $1/q^2$ series contains the condensates of the higher dimension. The condensates of the lowest dimension $d=3$ are the vector and scalar quark expectation values
$v(\rho)=\langle M|\sum_i\bar q^i\gamma_0q^i|M\rangle$ and
$\kappa(\rho)=\langle M|\sum_i\bar q^iq^i|M\rangle.$
Here $|M\rangle$ is the vector of state of nuclear matter, the sums are carried out over the quark flavors.
The vector condensate is model independent. It is just the density of the valence quarks.

Calculation of the scalar condensate requires certain assumptions  on the structure of the state $|M\rangle$.
In our earlier papers \cite{4},\cite{6},\cite{7} we started the analysis treating the nuclear matter as
a gas  of free noninteracting nucleons. In the present paper we follow another strategy. We assume the state $|M\rangle$ to be a system of interacting relativistic nucleons moving in self consistent scalar and vector fields. This system is placed in the QCD vacuum. In both approaches the scalar condensate is determined by the nucleon sigma term $\sigma_N$ which is related to observables. The main contribution with dimension $d=4$ is also related to observables. Determination of the contribution with $d=6$ dominated by the four quark condensates requires model assumptions on the quark structure of the nucleon.

We include consequently the condensates of higher dimension and the radiative corrections in calculation of nucleon parameters.
We find that after inclusion of the condensates with $d \leq 6$ the nuclear matter obtains equilibrium states for reasonable values of the sigma term $\sigma_N$.

Note that in the present analysis we include  only the contributions corresponding to the $2N$ interactions between the probe proton and the matter.
Each of QCD condensates is presented as the sum of its vacuum value and the expectation value of the same operator summed over the nucleons of the matter.

We give recall the main points of the finite density QCD sum rules approach in Sec.2
We present the expression for the binding energy in Sec.3
In Sec.~4-6 we calculate the nucleon parameters $m^*$ and $\Sigma_V$ and the energy per nucleon ${\cal B}$ including consequently the condensates of the higher dimension. We summarize in Sec.~7.

\section{Finite density QCD sum rules}

The vacuum sum rules \cite{2},\cite{3} are based on dispersion relations for the function $\Pi^{(0)}(q^2)=i\int d^4xe^{i(q\cdot x)} \langle 0|T[j(x)\bar j(0)]|0 \rangle,$ describing the propagation in vacuum of the system with the four-momentum $q$ which carries the quantum
numbers of the hadron (this is the proton in our case).
Here $j(x)$ is the local quark operator with the proton quantum numbers.
In similar way the finite density sum rules are based on dispersion relations
for the function $\Pi(q^2,s)=i\int d^4xe^{i(q\cdot x)} \langle M|T[j(x)\bar j(0)]|M\rangle$
describing the propagation of the system in nuclear matter.The functions $\Pi^{(0)}(q^2)$ and $\Pi(q^2,s)$ are often referred to as the "polarization operators"
in vacuum and in nuclear matter correspondingly.

In the dispersion relations for the three components $i=q,I,P$
\begin{equation}
\Pi^i(q^2,s)=\frac{1}{\pi}\int\frac{dk^2{\mbox{Im}}\Pi^i(k^2,s)}{k^2-q^2},
\label{NO}
\end{equation}
the left hand side is presented by several lowest terms of the OPE series (this will be labeled by the upper index $OPE$).
The imaginary parts on the right hand sides describe the physical states with the proton quantum numbers and are approximated by the "pole+continuum" model
in which the lowest lying pole is written down exactly,
while the higher states are described by continuum with the discontinuity corresponding to the OPE terms.
$$\frac{\Pi^i(k^2,s)}{\pi}=\zeta^i\lambda_{Nm}^2(s)\delta(k^2-m_m^2(s))+\frac{\Delta \Pi^{iOPE}(k^2,s)}{\pi}\theta(k^2-W_m^2(s)).$$
Here  $\zeta^q=1$, $\zeta^I=m^{*}$, $\zeta^P=-\Sigma_V/m$, while $m$ is the vacuum nucleon mass. The vector and scalar self-energies $\Sigma_V$ and $\Sigma_S=m^{*}-m$ and also the parameters $\lambda_{Nm}^2$ and $W_m^2$ are the unknowns of the sum rules equations. The position of the pole corresponding to the probe proton can be expressed in terms of $\Sigma_V$ and $\Sigma_S$; in linear approximation $m_m=m+\Sigma_S+\Sigma_V$.

The polarization operator $\Pi(q^2,s)$ can be presented as the sum of the vacuum term $\Pi^{0}(q^2)$ and the contribution $\Pi^{(m)}(q^2,s)$ caused by the nucleons of the matter, i.e.
$\Pi(q^2,s)=\Pi^{0}(q^2)+\Pi^{(m)}(q^2,s)$. There is no vacuum term for the $P$ structure, and  $\Pi^P(q^2,s)=\Pi^{P(m)}(q^2,s)$. Also the OPE expansion of the structures $\Pi^i(q^2,s)$ can be written as
$$\Pi^{q,OPE}(q^2,s)=\sum_{d=0}\pi^q_d(q^2,s); \quad \Pi^{I,OPE}(q^2,s)=\sum_{d=3}\pi^I_d(q^2,s);$$
$$\quad \Pi^{P,OPE}(q^2,s)=\sum_{d=3}\pi^p_d(q^2,s),$$
with $d$ the dimensions of the condensates. The lowest terms for the OPE expansions of the $I$ and $q$ structures contain the condensates with dimension $d=3$. The term with $d=0$ in the OPE expansion of $\Pi^{q,OPE}$ is the free three-quarks loop. The calculations are carried out in the chiral limit, with the light quark masses $m_{u,d}=0$.

In next step the Borel transform (the inverse Laplace transform) is used. Operator $\hat B=(Q^2)^{n+1}/n!\times (-\partial/\partial Q^2)^n$
with $Q^2=-q^2$, $Q^2,n \rightarrow \infty$ converts a function $f(q^2)$ to $\hat Bf(q^2)=\hat f(M^2)$ depending on $M^2=Q^2/n$.
The Borel transform eliminates the divergent terms (thus we should not worry about subtractions in the dispersion relations). Also the Borel transform suppress the contribution of continuum in the dispersion relations.
Calculating $\hat B (k^2-q^2)^{-1}=e^{-k^2/M^2}$ and applying the operator $32\pi^4\hat B$ to both sides of Eq.(\ref{NO}) we obtain the
Borel transformed finite density QCD sum rules
\begin{equation}
{\cal L}^q(M^2, W_m^2;\eta)=\Lambda_m(M^2); \quad {\cal L}^I(M^2, W_m^2;\eta)=m^{*}\Lambda_m(M^2);
\label{X1}
\end{equation}
$${\cal L}^P(M^2, W_m^2;\eta)=-\frac{\Sigma_V}{m}\Lambda_m(M^2).$$
Here $\eta(\rho)$ is the set of the QCD condensates in nuclear matter with $\rho$ the baryon density of the matter.
The right hand sides describe the pole corresponding to the probe proton, $\Lambda_m=\lambda^2_{m}e^{-m_m^2/M^2}$ with $\lambda_m^2=32\pi^4\lambda_{Nm}^2$ while $\lambda_{Nm}^2$ is the residue of the nucleon pole. The factor $32\pi^4$ is introduced to deal with the quantities of the order of unity in powers of GeV.
Here $\eta(\rho)$ is the set of the QCD condensates in nuclear matter.

One can write
\begin{equation}
{\cal L}^q(M^2, W_m^2;\eta)=\sum_{d=0}^{\infty}A_d(M^2,W_m^2;\eta); \quad {\cal L}^I(M^2, W_m^2;\eta)=\sum_{d=3}^{\infty}B_d(M^2, W_m^2;\eta);
\label{X2}
\end{equation}
$${\cal L}^P(M^2, W_m^2;\eta)= \sum_{d=3}^{\infty}C_d(M^2,W_m^2; \eta),$$
with $d$ the dimension of the condensates, $A_d, B_d, C_d$ the Borel transformed OPE terms. The OPE terms of the dimensions $ d \leq 6$  are shown in Fig.~1. Recall that $A_0$ stands for the free three quark loop.

Each condensate can be presented as
\begin{equation}
\eta(\rho)=\eta^{(0)}+\eta^{(m)}(\rho),
 \label{X2a}
 \end{equation}
 with $\eta^{(0)}=\eta(0)$ the vacuum value of $\eta$, while $\eta^m(\rho)$ is caused by the nucleons of the matter. Some of the condensates obtain nonzero values only for $\rho \neq 0$, i.e. $\eta^{(0)}=0$. Also, the terms on the right hand sides of Eq.(\ref{X2}) can be written as
$$A_d(M^2, W_m^2; \eta(\rho))=A_d^{(0)}(M^2, W_m^2; \eta^{(0)})+A^{(m)}_d(M^2, W_m^2; \eta(\rho));$$
\begin{equation}
\quad B_d(M^2, W_m^2; \eta(\rho))=B_d^{(0)}(M^2, W_m^2; \eta^{(0)})+B^{(m)}_d(M^2, W_m^2; \eta(\rho));
\label{11a}
\end{equation}
$$C_d(M^2, W_m^2; \eta(\rho))=C^{(m)}_d(M^2, W_m^2; \eta(\rho)).$$
Here $A_d^{(0)}$ and  $B_d^{(0)}$ are the vacuum values, $C_d^{(0)}=0$. $A_d^{(m)},B_d^{(m)},C_d^{(m)}$ are due to exchange by quarks and the gluons between our
three quark system and the nucleons of the matter. They can depend on both $\eta^{(0)}$ and $\eta^{(m)}$. Also the left hand sides of Eq.({\ref{X1})
can be written as the sum of the vacuum contribution and that caused by the nucleons of the matter
$${\cal L}^i={\cal L}^{i(0)}+{\cal L}^{i(m)}.$$

The vacuum Borel transformed  sum rules \cite{3} are
\begin{equation}
{\cal L}^{q(0)}=\Lambda(M^2);\quad {\cal L}^{I(0)}=m\Lambda(M^2).
\label{11b}
\end{equation}
with
\begin{equation}
{\cal L}^{q(0)}=A^{(0)}_0(M^2, W^2)+  A^{(0)}_4(M^2, W^2)+A^{(0)}_6(M^2)=\Lambda(M^2);
\label{11bc}
\end{equation}
$${\cal L}^{I(0)}=B^{(0)}_3(M^2, W^2)+B^{(0)}_7(M^2, W^2).$$
Here $W_0^2$ is the vacuum value of the threshold.
Also  $\Lambda=\lambda_0^2e^{-m^2/M^2}$ with $\lambda_0^2$ the vacuum value of the residue in the proton pole.
The leading logarithmic radiative corrections $\alpha_s\ln{M^2}$ are summed in all orders.
The radiative corrections of the order $\alpha_s$ are included in the first order of perturbative theory.
The terms on the right hand sides of Eq.(\ref{11bc}) are
 \begin{equation}
A^{(0)}_0=\frac{M^6E_2(W^2/M^2)}{L^{4/9}}\Big(1+\frac{71}{12}\frac{\alpha_s}{\pi}\Big); \quad A^{(0)}_4=\frac{bM^2E_0(W^2/M^2)}{4};
\label{11x}
\end{equation}
$$A^{(0)}_6(M^2)=\frac{4}{3}a^2\Big(1-\frac{5}{6}\frac{\alpha_s}{\pi}\Big); \quad B_3=2aM^2E_1(W^2/M^2)\Big(1+\frac{3}{2}\frac{\alpha_s}{\pi}\Big); \quad
B_7=-\frac{ab}{12}.$$
Here the functions $E_0(x)=1-e^{-x}$, $E_1(x)=1-(1+x)e^{-x}$ and  $E_2(x)=1-(1+x+x^2/2)e^{-x}$ describe the contributions of continuum which approximates the higher excited states. The factors $a$ and $b$ are proportional to the quark and gluon vacuum condensates: $a=-(2\pi)^2\langle 0|\bar uu|0\rangle $, $b=(2\pi)^2
\langle 0|(\alpha_s/\pi)G^{a\mu\nu}G_{a\mu\nu}|0\rangle$. The conventional values are $\langle 0|\bar uu|0\rangle =(-240 MeV)^3$ and
$\langle 0|(\alpha_s/\pi)G^{a\mu\nu}G_{a\mu\nu}|0\rangle=(0.33 GeV)^4$ \cite{3}. The function $L=(\ln{M^2/\Lambda_{QCD}^2})/(\ln{\mu^2/\Lambda_{QCD}^2})$ with $\Lambda _{QCD}=230$ MeV accounts for the leading logarithmic radiative corrections. The standard normalization point is $\mu=500$ MeV.

The solution is found by minimization of the difference between left hand sides and right hand sides of Eq.(\ref{11x}) in
the interval \cite{3}
\begin{equation}
0.8 GeV^2 \leq M^2 \leq 1.4 GeV^2.
\label{12N}
\end{equation}
This provides $m=928$~MeV for the nucleon mass \cite{3a}.
We employ this value in further calculations.
Also $\lambda_0^2=2.36$ GeV $^6$ and $W_0^2=2.13$ GeV $^2$ are the vacuum values of the nucleon residue and of the continuum threshold correspondingly.
We shall add consequently the terms $A_d^{(m)}$, $B_d^{(m)}$ and $C_d^{(m)}$ to Eq.~(\ref{11b}).

Employing Eq.~(\ref{X1}) one can present the effective mass of the probe nucleon $m^*$ and the vector self energy $\Sigma_V$  as
\begin{equation}
\frac{{\cal L}^I(M^2, W_m^2; \eta(\rho))}{{\cal L}^q(M^2, W_m^2;\eta(\rho))}=m^{*};\quad
\frac{{\cal L}^P(M^2, W_m^2; \eta(\rho))}{{\cal L}^q(M^2, W_m^2; \eta(\rho))}=-\Sigma_V/m.
\label{12}
\end{equation}
The unknown $W_m^2$ ties the two equations. Neglecting the in-medium shift of the vacuum threshold $W_0^2$, i.e. putting $W_m^2=W_0^2$ we find two independent expressions for the nucleon parameters. One can see that $m^*$ and $\Sigma_V$ do not exhibit linear dependence on $\rho$ even if we include only the  linear $\rho$ dependence of the condensates.

\section {Binding energy}

The matter has an equilibrium bound state at $\rho=\rho_{eq}$ if the average binding energy per nucleon
\begin{equation}
{\cal B}(\rho)=\frac{{\cal E}(\rho)}{\rho}-m
\label{14}
\end{equation}
reaches its minimum value at this point, i.e. $d{\cal B}/{d\rho}|_{\rho=\rho_{eq}}=0$, and  ${\cal B}(\rho_{eq})<0$.
The phenomenological equilibrium value is
$$ \rho_0=0.17{\mbox { fm}}^{-3},$$
with ${\cal B}(\rho_0)\approx -16$~MeV (see e.g.\cite{Xa}).

To calculate ${\cal E}$ in terms of $m^*$ (or $\Sigma_S=m^*-m$) and $\Sigma_V$ we write the wave equation for the nucleon with the four-momentum $p$.
$$ \Big(\hat p-\frac{\hat P}{m}\Sigma_V+m^*\Big)u=0; \quad P=(m, {\bf 0}).$$
The bispinor $u$ is normalized by condition $\bar u\gamma_0u=1$.
The energy of the nucleon is
\begin{equation}
p_0=E(|{\bf p}|)=\bar u(p_i\gamma^i+m^*)u+\Sigma_V\bar u\gamma_0u,
\label{15}
\end{equation}
with $m^*$ and $\Sigma_V$ the effective mass and the vector self energy which can be found from Eq.~(\ref{12})(The probe nucleon has the three- momentum $|{\bf p}|=p_F$).
Note that in nuclear mater the self energies $\Sigma_S$ and $\Sigma_V$ do not depend on ${\bf p}$.

The three momenta of the nucleons $p$ can not exceed the Fermi momentum $p_F$. The latter is connected with baryon density $\rho$ by the relation
\begin{equation}
\rho=4\int\frac{d^3p}{(2\pi)^3}\theta(p_F-p)=\frac{2}{\pi^2}\int_{0}^{p_F}dpp^2=\frac{2 p_F^3}{3\pi^2}.
\label{4a}
\end{equation}
If only the $2N$ forces are included \cite{14},
\begin{equation}
{\cal E}=4\int \frac{d^3p}{(2\pi)^3}\theta (p_F-p)\Phi(p); \quad \Phi(p)=\bar u(p_i\gamma^i+m)u+\frac{\Sigma_V}{2}\bar u\gamma_0u+\frac{\Sigma_S}{2}\bar uu; \quad \Sigma_S=m^*-m.
\label{16}
\end{equation}

Employing the equalities $\bar uu=m^*/E^*$ and $\bar u p_i\gamma^iu=p^2/E^*$ with  $E^*=\sqrt{m^{*2}+p^2}$ \cite{10}.
one can find
\begin{equation}
{\cal E}=4\int \frac{d^3p}{(2\pi)^3}\theta (p_F-p)E^*(p)-\frac{\Sigma_S}{2}\rho F(\rho)+\frac{\Sigma_V}{2}\rho.
\label{17}
\end{equation}

Here
\begin{equation}
F(\rho)=\frac{4}{\rho}\int\frac{d^3p}{(2\pi)^3}\frac{\theta(p_F-p)}{\gamma(p)}=
\frac{2}{\pi^2\rho}\int_{0}^{p_F}dpp^2\frac{1}{\gamma(p)},
\label{7}
\end{equation}
with
$$ \gamma(p)=\frac{\sqrt{m^{*2}+p^2}}{m^*}=\frac{1}{\sqrt{1-v^2}},$$
the Lorenz factor of the nucleon moving with the velocity  $v=
p/m^*$.

This enables us to calculate ${\cal B}(\rho)$ following the definition (\ref{14}).
In the nonrelativistic limit $ p_F^2 \ll m^{*2}$ we come to the well known expression
\begin{equation}
{\cal B}(\rho)=\frac{3}{5}\frac {p_F^2}{2m^*(\rho)}+\frac{\Sigma_V(\rho)}{2}+ \frac{\Sigma_S(\rho)}{2},
\label{17a}
\end{equation}
with $\Sigma_S(\rho)=m^*(\rho)-m$.

\section{Contributions of the condensates of lowest dimension caused by the nucleons of the matter}

Here we include only the contributions of condensates with $d=3$.
These are the vector and the scalar condensates.
The vector condensate written in the rest frame of the matter is
\begin{equation}
v(\rho)=\langle M|\sum_i\bar q^i(0)\gamma_0q^i(0)|M\rangle= n_v\rho,
\label{X3}
\end{equation}
with $n_v=3$ the number of the valence quarks in nucleon. Thus the vector condensate is exactly proportional to the nucleon density.
The scalar condensate is
\begin{equation}
\kappa(\rho)=\langle M|\sum_i\bar q^i(0)q^i(0)|M\rangle.
\label{X100}
\end{equation}

Thus we included exchange by quark-antiquark pairs between out three-quark system and the nucleons of the matter in the vector and scalar channels.
Only the quark interactions at very small distances of the order of $1 GeV^{-1}$ are included. They are described by the factor $L=(\ln{M^2/\Lambda_{QCD}^2})/(\ln{\mu^2/\Lambda_{QCD}^2})$-see the text after Eq.(\ref{11x}). Note that absorbtion and radiation of quarks takes place at the same space-time point.

The $\hat q$ and $\hat P$ structures $A_3$ and $C_3$ are proportional to the vector condensate. Since $v(\rho)=0$ at $\rho=0$, $A_3^{(0)}=C_3^{(0)}=0$ and
$A_3=A_3^{(m)}$ while $C_3=C_3^{(m)}$. The $I$ structure $B_3$ is proportional to the scalar condensate with $B_3^{(0)}$ contributing to the vacuum sum rules.
The left hand sides of Eq.(\ref{X1}) can be presented as
\begin{equation}
{\cal L}^q= {\cal L}^{q(0)}+A_3; \quad {\cal L}^I= {\cal L}^{I(0)}+B_3^{(m)}; {\cal L}^P= C_3.
\label{17z}
\end{equation}

Direct calculation provides
\begin{equation}
A_3=-\frac{8\pi^2}{3}\frac{(s-m^2)M^2E_0(W_m^2/M^2)-M^4E_1(W_m^2/M^2)}{mL^{4/9}}v(\rho);
\label{18}
\end{equation}
$$B_3^{(m)}=-4\pi^2M^4E_1(W_m^2/M^2 )\Big(\kappa(\rho)-\kappa(0)\Big); \quad C_3=-\frac{32\pi^2}{3}\frac{M^4E_1(W_m^2/M^2)}{L^{4/9}}v(\rho).$$
As in Eq.(\ref{11x}), the functions $E_0(x)$ and $E_1(x)$ describe the contributions of the higher excited states.

While the vector condensate can be obtained in model independent way, calculation of the scalar condensate requires certain assumptions  on the structure of the state $|M\rangle$.
We assume the state $|M\rangle$ to be a system of interacting relativistic nucleons moving in self consistent scalar and vector field. The system is placed into the QCD vacuum.
Thus we can write
\begin{equation}
\kappa(\rho)=\kappa(0)+\kappa^{(m)}(\rho); \quad \kappa(0)=\langle 0|\sum_i\bar q^iq^i|0\rangle ; \quad \kappa^{(m)}(\rho)=\langle \tilde M|\sum_i\bar q^iq^i|\tilde M\rangle.
\label{X101}
\end{equation}
with  $|\tilde M\rangle$ the vector of state describing the nucleons of the matter.
The first term on the right hand side is the vacuum value
$\kappa(0)=\langle 0|\bar uu+\bar dd|0\rangle$. Neglecting the small $SU(2)$ breaking effects we can put $\kappa(0)=2\langle 0|\bar uu|0\rangle =2\langle 0|\bar dd|0\rangle$. The conventional value is $\kappa(0)=2(-240 MeV)^3$ \cite{3}.

The second term on the right hand side of Eq.(\ref{X101}) can be presented as
\begin{equation}
\kappa^{(m)}(\rho)=\langle \tilde M|\sum_i\bar q^iq^i|\tilde M\rangle=\langle\tilde M|\frac{\partial H}{\partial m_q}|\tilde M\rangle ; \quad m_q=\frac{m_u+m_d}{2},
\label{X4}
\end{equation}
with $H(x)=m_u\bar u(x)u(x)+m_d\bar d(x)d(x)+...$ the
density of the QCD Hamiltonian.  Here the dots stand for the terms which do not depend on the
masses of the light quarks $m_u$ and $m_d$. Note that $\langle\tilde M|H|\tilde M\rangle={\cal E}$. The latter is determined by Eq.(\ref{17}).
The Hellmann--Feynman theorem \cite{6a} enables to evaluate the second term on the right hand side of Eq.(\ref{X4})
$$\langle \tilde M|\frac {\partial H}{\partial m_q}|\tilde M\rangle=\frac {\partial {\cal E}}{\partial m_q},$$
(the derivatives of the vectors of state $\langle M|$ and $|M\rangle$ cancel). Hence
\begin{equation}
\kappa(\rho)=\kappa(0)+\frac{\partial {\cal E}}{\partial m_q}.
\label{X05}
\end{equation}

The energy density ${\cal E}$ determined by Eq.(\ref{17})depends on nucleon characteristics $m$, $\Sigma_S$ and $\Sigma_V$, depending on the quark mass $m_q$. The derivative $\partial m/\partial m_q$ can be expressed through the nuclear sigma term
$\sigma_N$ related to the low energy pion-nucleon scattering amplitude \cite{8}
\begin{equation}
\sigma_N=m_q\frac{\partial m}{\partial m_q}.
\label{X60}
\end{equation}
The conventional value is $\sigma_N=(45\pm 8)$~MeV \cite{8}. This provides
$$\frac{\partial m}{\partial m_q}=\frac{\sigma_N}{m_q} \approx 8.$$
Note than employing the Hellmann--Feynman theorem for the free nucleon at rest described by the vector of state $|N\rangle$ one finds that
\begin{equation}
\frac{\partial m}{\partial m_q}=\langle N|\sum_i\bar q^iq^i|N\rangle=\kappa_N,
\label{X61}
\end{equation}
is the expectation value of the operator $\sum_i\bar q^iq^i$ in free nucleon.

Our assumption is that the main dependence of the energy density ${\cal E}$ on the quark mass $m_q$ is contained in the nucleon mass $m$.
It is based on the estimation of the quark mass dependence of contribution provided by diagrams shown  in Fig.~1a,b made in \cite {9a}. The estimations show that  $(m/\Pi_i)\partial\Pi_i/\partial m_{u,d} \ll \kappa_N$, and such contributions can be neglected. Also, the assumption is supported by relatively small
contribution of the nucleon interactions  to the condensate $\kappa(\rho)$ found in numerous approaches \cite{4},\cite{9}.

Thus, neglecting the derivatives $\partial \Sigma_S/\partial m_q$ and $\partial \Sigma_V/\partial m_q$ we obtain
\begin{equation}
\kappa(\rho)=\kappa(0)+\kappa^{eff}_N(\rho)\rho,
\label{8}
\end{equation}
with
\begin{equation}
\kappa_N^{eff}(\rho)=\kappa_NF(\rho),
\label{9}
\end{equation}
while $F(\rho)$ is defined by Eq.(\ref{7}). Thus the function $F(\rho)$ describes the modification of the nucleon expectation value $\kappa_N$ caused be relativistic motion of the
nucleons composing the matter.

The function $F(\rho)$ also ties the baryon and scalar densities in the mean field solution of the
scalar-vector model of nuclear matter, known also as the Walecka model \cite{10}. Note that $F(\rho)=1$ in the nonrelativistic limit $p_F \ll m^*$.
Since the function $F(\rho)$ depends explicitly on the nucleon effective mass $m^*$, we shall write it sometimes as $F(\rho, m^*(\rho))$ employing
also the notations $\kappa_N^{eff}(\rho, m^*(\rho))$ and
 $\kappa(\rho, m^*(\rho))$.
 Hence expression for $B_3^{(m)}$ in Eq.(\ref{18}) can be written as
 \begin{equation}
B_3^{(m)}=-4\pi^2M^4E_1(W_m^2/M^2 )\kappa_N^{eff}(\rho)\rho,
 \label{18s}
\end{equation}
 with $\kappa_N^{eff}$ defined by Eq.(\ref{9}).

Including only the condensates of the lowest dimension $d \leq 3$ and employing Eq.~(\ref{X2}) we write Eq.~(\ref{X1}) as
\begin{equation}
{\cal L}^{q(0)}(M^2, W_m^2)+A_3(M^2, W_m^2;v(\rho))=\Lambda_m(M^2);
\label{10}
\end{equation}
$${\cal L}^{I(0)}+B_3^{(m)}(M^2,W_m^2;\kappa(\rho, m^*(\rho))=m^{*}(\rho)\Lambda_m(M^2);$$
$$C_3(M^2;v(\rho))=-\frac{\Sigma_V(\rho)}{m}\Lambda_m(M^2).$$
The vacuum terms ${\cal L}^{q(0)}$ and ${\cal L}^{I(0)}$ are given by Eq. (\ref{11bc}). The first and the last equations just tie the nucleon parameters with the vector condensate $v(\rho$). The second one contains dependence on $m^*(\rho)$ in both sides. Thus, solving Eqs.~(\ref{10}) we find the nucleon parameters and the in-medium scalar condensate
$\kappa(\rho)$ self consistently.
The values of nucleon parameters depend on the value of nucleon sigma term $\sigma_N$. The conventional value is $\sigma_N=(45\pm 8)$~MeV \cite{8}.
However some recent results \cite{11} are consistent with smaller values, i.e. $\sigma_N=(44\pm 12)$~MeV.
Also, one can meet the larger values of $\sigma_N$ in literature (see, e.g. \cite{12}). The value $\sigma_N=(66\pm 6)$~MeV \cite{13} is the largest one. We consider the values of the sigma term in the interval
\begin{equation}
35 {\mbox { MeV}} \leq \sigma_N \leq 65 {\mbox { MeV}}.
\label{11}
\end{equation}

One can make a rough estimation for the values of the nucleon self energies. Assuming that the vacuum sum rules equations given by Eq.(\ref{11b})
hold exactly and putting $W_m^2=W^2$ we write Eq.(\ref{12}) as
\begin{equation}
m^{*}=\frac{m+h^I(M^2)}{1+h^q(M^2)}; \quad \Sigma_V=-\frac{h^P(M^2)}{1+h^q(M^2)}.
\label{c2x}
\end{equation}
We denoted
$h^q(M^2)=A_3(M^2)/\Lambda(M^2)$, $h^I(M^2)=B^{(m)}_3(M^2)/\Lambda(M^2)$,
 and $h^P(M^2)=C_3(M^2)/\Lambda(M^2)$.
Recall that $\Lambda(M^2)=\lambda_0^2e^{-m^2/M^2}$.
Introducing the functions
$$ f^q(M^2)=\frac{(s-m^2)M^2E_0(W_0^2/M^2)-M^4E_1(W_0^2/M^2)}{mL^{4/9}}\frac{e^{m^2/M^2}}{\lambda_0^2},$$
$$f^I(M^2)=M^4E_1(W_0^2/M^2)\frac{e^{m^2/M^2}}{\lambda_0^2},$$
$$ f^P(M^2)=\frac{M^4E_1(W_0^2/M^2)}{L^{4/9}}\frac{e^{m^2/M^2}}{\lambda_0^2},$$
we can write the ingredients of Eq.(\ref{c2x}) as
\begin{equation}
h^q(M^2)=-\frac{8\pi^2}{3}f^q(M^2)v(\rho); \quad h^I(M^2)=-4\pi^2f^I(M^2)\kappa_N^{eff}(\rho);
\label{cx1}
\end{equation}
$$h^P(M^2)=-\frac{32\pi^2}{3}f^I(M^2)v(\rho).$$

The functions $f^i(M^2)$ ($i=q,I,P$) vary very slowly in the interval determined by Eq.(\ref{12N}). There values change by about 10 percent in this interval. This enables us to replace the functions $f^i(M^2)$ ($i=q,I,P$) by
their average values $\beta^q=1.28GeV^{-3}$,  $\beta^I=0.65GeV^{-2}$, and $\beta^P=0.49GeV^{-2}$.

Thus the nucleon parameters can be presented as
\begin{equation}
m^{*}=\frac{m+{\cal F}^I}{1+{\cal F}^q}; \quad \Sigma_V=-\frac{{\cal F}^P}{1+{\cal F}^q},
\label{cx}
\end{equation}
where the parameters ${\cal F}^i$ are the averaged values of the functions $h^i(M^2)$.

Employing these values we find that ${\cal F}^q=-0.13$, ${\cal F}^P=-204$ MeV for $\rho=\rho_0$. Thus $\Sigma_V \approx 235 $ MeV.
Similar estimation for the effective mass can be obtained by neglecting the influence of the Lorentz factor on the scalar condensate, i.e. by putting
$\kappa_N^{eff}=\kappa_N$ in Eq.(\ref{18s}) for $B_3$. For the value of $\sigma_N$ on the upper limit of the interval determined by Eq.(\ref{11}),
i.e. at $\sigma_N=65$ MeV, with $\kappa_N=11.8$, we find ${\cal F}^I \approx-400$ MeV for $\rho=\rho_0$, providing $m^* \approx 607$ MeV, and $\Sigma_S \approx -321$ MeV.

Solutions of Eq.(10) (with $F(\rho)=1$) provide $\Sigma_V=293$ MeV and $m^*=569$ MeV for $\rho=\rho_0$ and $\sigma_N=65$ MeV.
Solving Eq.(10) with inclusion of the factor $F(\rho)$ we find $\Sigma_V=295$ MeV and $m^*=601$ MeV ($\Sigma_S=-327$ MeV) at this point.

Now we calculate the binding energy. The result is shown in Fig 2. The function ${\cal B}(\rho)$ obtains its minimal value at a density close to $\rho_0$.
However ${\cal B}(\rho)$ is always positive. Thus there are no equilibrium states.
This is true for the smaller values of the sigma term $\sigma_N$. Indeed, one can see from Eq.(\ref{17}) that
$\partial{\cal B}(\rho)/\partial m^*>0$. On the other hand $m^*$ drops with $\sigma_N$. Hence $\partial{\cal B}(\rho)/\partial \sigma_N<0$,
and ${\cal B}(\rho)$ is larger for smaller values of $\sigma_N$.

Note that the shape of the curve ${\cal B}(\rho)$ is due to the influence of the relativistic motion of nucleons on the scalar quark condensate.
Indeed, putting $F=1$ in Eq.(\ref{9}) (but keeping it on the right hand side of Eq.(\ref{17})), and thus assuming $\kappa_N^{eff}=\kappa_N$ we find the monotonically dropping curve
shown by dashed line in Fig.2.

\section{Inclusion of radiative corrections and of next to leading OPE terms}

 Now we include the leading radiative corrections to the contributions $A_3^{(m)}$, $B_3^{(m)}$ and $C_3^{(m)}$ beyond the logarithmic approximation.
 This means that we include the corrections of the order $\alpha_s$ which do not contain large logarithmic factors $\ln {M^2}$. The latter have been
 included in calculations carried out in the previous Section. This means that the quark-antiquark pairs which are exchanged between our three-quark system and the matter experience interactions at the distances of the order $\Lambda_{QCD}^{-1}$. These interactions are rather weak and are included in the lowest order of perturbative theory. The corresponding contributions are \cite{4}
\begin{equation}
A_3^{r}=A_3\cdot\frac{7}{2}\frac{\alpha_s}{\pi}; \quad B_3^{(m)r}=B_3^{(m)}\cdot\frac{3}{2}\frac{\alpha_s}{\pi};\quad
C_3^{r}=C_3\cdot\frac{15}{4}\frac{\alpha_s}{\pi}.
\label{18an}
\end{equation}
We employ the value $\alpha_s=0.475$ corresponding to $\Lambda_{QCD}=230 MeV$.

We include also the contributions of condensates with $d=4$. One of them is caused by the gluon condensate.
The gluon condensate provides the term $A_4^{(0)}$ in the vacuum sum rule for the $\hat q$ structure given by Eq.({\ref{11b}) describing the exchange by gluons between the free three-quark loop and the QCD vacuum. The gluon exchange between the three quark loop and the nucleons of the matter is given by the term
\begin{equation}
A_{4a}^{(m)}=\pi^2M^2E_0(W_m^2/M^2)g(\rho),
\label{18a}
\end{equation}
with $g(\rho)=g_N\rho$, $g_N=\langle N|(\alpha_s/\pi)G^{a\mu\nu}G_{a\mu\nu}| N\rangle$, with $| N\rangle$ describing the nucleon in the nuclear matter.
$G^{\mu\nu}$ is the tensor of the gluon field, $a$ is the color index -- see Fig.~1$d$.

Another type of contributions comes from the nonlocality of the vector condensate \newline
$\langle M|\bar q(0)\gamma_0q(x)|M\rangle$. To be consistent with the gauge invariance one should treat the operator $q(x)$ as
the Taylor series $q(x)=(1+x_{\alpha}D_{\alpha}+1/2\cdot x_{\alpha}x_{\beta}D_{\alpha}D_{\beta}+...)q(0)$ providing infinite series of the local condensates. Only the second term in parenthesis leads to the condensates of dimension $d=4$.
Numerically most important ones can be expressed in terms of the moments ${\cal M}_n=\int_0^1d\alpha\alpha ^{n-1}f(\alpha)$ of the nucleon structure function normalized by condition ${\cal M}_1=\int_0^1d\alpha f(\alpha)=n_q$ ($n_q=3$ is the number of valence quarks in a nucleon). The most important contribution is provided by the second moment ${\cal M}_2$. Its contribution to the $\hat q$ structure is
\begin{equation}
A_{4b}^{(m)}=\frac{16\pi^2}{3}mM^2E_0(W_m^2/M^2){\cal M}_2\rho.
\label{19}
\end{equation}
Its combination with that of the gluon condensate provides the contribution to the $q$ structure
\begin{equation}
  A_{4}^{(m)}=A_{4a}^{(m)}+A_{4b}^{(m)}=\pi^2 M^2E_0(W_m^2/M^2)\Big (g_N+\frac{16}{3}{\cal M}_2\Big)\rho.
\label{19N}
\end{equation}
The contribution to the $P$ structure is
\begin{equation}
C_4^{(m)}=\frac{8\pi^2}{3}5\frac{(s-m^2)M^2E_0(W_m^2/M^2)-M^4E_1(W_m^2/M^2)}{m}{\cal M}_2\rho.
\label{19a}
\end{equation}

Note that the term $B_4^{(m)}$ corresponding to nonlocality of the scalar condensate  vanishes in the chiral limit.
The contribution $B_4^{(m)}$ is proportional to the expectation value $\langle M|\bar q(0)D_{\mu}q(x)|M\rangle$ with $q=u,d$ denoting a light quark.
Presenting $D_{\mu}=(\gamma_{\mu}\hat D+\hat D\gamma_{\mu})/2$ and employing the QCD equation of motion $\hat Dq=m_qq$ we find that $B_4^{(m)}$
is indeed proportional to $m_q$. Thus in out approach $B_4^{(m)}=0$.

The moments of the nucleon structure functions are not changed noticeably in nuclear medium. Also, the contribution of the gluon condensate
is numerically small. Thus we employ the values of the parameters of free nucleons ${\cal M}_2=0.32$ \cite{16}, and $g_N=-8/9m$ \cite{4}.

The rough estimation can be made in the same way as in previous Section. We present the estimations for $\rho=\rho_0$. The radiative corrections and those caused by nonlocality of the vector condensate add $-0.07$ and $0.01$ correspondingly to the value of ${\cal F}^q$. Thus the new value is ${\cal F}^q=-0.19$. The radiative corrections diminish ${\cal F}^P$ by $116$ MeV. However this is compensated to large extent by the contribution presented by Eq.(\ref{19a}) which increases ${\cal F}^P$ by $95$ MeV. Thus the vector self energy $\Sigma_V \approx 280$ MeV. The radiative corrections to the scalar structure ${\cal F}^I$ diminish it  by $90$ MeV for $\sigma_N=65$ MeV and $\kappa_N^{eff}=\kappa_N$. This provides $m^* \approx 540$ MeV
Thus we can expect that the effects considered in this Section increase the value of $\Sigma_V$ by about $40$ MeV and diminish the value of $m^*$
by about $70$ MeV at $\rho=\rho_0$ and $\sigma_N=65$ MeV. Note. however that inclusion of the effects discussed in this Subsection lead to noticeable modifications of the values of the continuum threshold $W_m^2$- see Table 1 below. This makes the numbers obtained by such estimation less reliable.

Solving Eqs.~({\ref{X1}) we find that the values of nucleon parameters are consistent with the estimated ones. Putting $\kappa_N^{eff}=\kappa_N=11.8$
corresponding the $\sigma_N=65$ MeV we find $\Sigma_V=333$ MeV and $m^*=460$ MeV, at $\rho=\rho_0$. Employing Eq.(\ref{9}) for the value
of $\kappa_N^{eff}$ we obtain $\Sigma_V=336$ MeV and $m^*=506$ MeV, $\Sigma_S=-422$~MeV at this point. These values are rather close to those of the Walecka model where
the equilibrium is reached ar $\rho_{eq}=1.13\rho_0$. At this point $\Sigma_V=323$~MeV, while $\Sigma_S=-413$~MeV. Thus we can expect that there are equilibrium states of the matter for the values of $\sigma_N$ close to the upper limit $\sigma_N=65$ MeV and at $\rho$ around the point $\rho_0$.

Direct calculations demonstrate that energy ${\cal B}$ determined by Eqs.(\ref{14},\ref{17}) obtains a minimum for $\sigma_N \geq 60$~MeV -- see Fig.~3.
The value of $\rho_{eq}$ is very close to the phenomenological equilibrium value $\rho_0$ for $\sigma_N \approx 65$~MeV. We find $\rho_{eq}=0.95\rho_0$ for $\sigma_N=64.5$~MeV with ${\cal B}_{min}=-15.8$~MeV.
Note that the existence of equilibrium states
is due to the influence of the relativistic motion of nucleons on the scalar quark condensate.
Putting $F(\rho)=1$ in Eq.(\ref{cx}) for $m^*$ we see that the numerator exhibits the linear drop with density. Since ${\cal F}^q<0$, the effective mass drops even faster. This leads to monotonic dropping curve for ${\cal B}(\rho)$-see the dashed line in Fig 3.
Inclusion of the factor $F(\rho)$ slowers the drop of $m^*$

In Fig.~4 we present the results for nucleon parameters for $\sigma_N=65$~MeV with  $\rho_{eq}=0.99\rho_0$ and ${\cal B}_{min}=-17.8$~MeV. In Fig.~4$a$ we show the density dependence of the vector self energy
$\Sigma_V$ and for the nucleon effective mass $m^*$. In Fig.~4$b$ we demonstrate the density dependence of the residue  $\lambda_m^2$ and that of the effective threshold $W_m^2$. The function $F$ is determined by Eq.~(\ref{7}). More detailed results for the values of density close to $\rho_{eq}$ are presented in Table 1. At the equilibrium point $\Sigma_S=-418$~MeV while $\Sigma_V=334$~MeV.
Note that the value of the nonrelativistic single particle potential energy $U=\Sigma_S+\Sigma_V=-84$ MeV is close to that of the Walecka model
$U=-90$ MeV at its equilibrium point
In Fig.~5 we present dependence of ${\cal B}_{min}$ on the value of $\sigma_N$. The dependence of $\rho_{eq}$ on $\sigma_N$ is shown in
Fig.~6.

\begin{table}
\caption{Nucleon parameters obtained with inclusion of the condensates with $d \leq 4$ and of the lowest order radiative corrections,
$\sigma_N=65$~MeV.
For each value of the baryon density $\rho$ the upper line corresponds to $\kappa_N^{eff}$ determined by Eq.(\ref{9}).
The lower line is for $\kappa_N^{eff}=\kappa_N$.}

\begin{center}
\begin{tabular}{|c|c|c|c|c|c|}
\hline
$\rho/\rho_0$&$m^*$,~MeV&$\Sigma_V$,~MeV&$\lambda_m^2$,~GeV$^6$&$W_m^2$,~GeV$^2$\\
\hline
0.90&552&294&1.55&1.84\\
&519&292&1.46&1.78\\
\hline
0.95&529&314&1.53&1.83\\
&490&312&1.42&1.76\\
\hline
1.00&506&336&1.50&1.83\\
&460&333&1.39&1.75\\
\hline
1.05&484&358&1.49&1.83\\
&428&355&1.35&1.73\\
\hline
1.10&462&380&1.47&1.82\\
&396&377&1.32&1.72\\
\hline
\end{tabular}
\end{center}
\end{table}

\section{Inclusion of the four-quark condensates}

Now we include the terms with dimension $d=6$. These are the four-quark condensates
described by the matrix elements
$\langle M|q^{ia}_{\alpha}\bar q^{ib}_{\beta}q^{jc}_{\gamma}\bar q^{jd}_{\delta}|M\rangle$ with $i=u,d$ and $j=u,d$. For each of the products of the quark operators (here we omit the flavor indices) we can write
$$
q^a_{\alpha}\bar q^b_{\beta}=-\frac{1}{12}\sum_X\bar q\Gamma_Xq(\Gamma^X)_{\alpha\beta}\delta_{ba}-
\frac{1}{8}\sum_{X,a}\bar q\Gamma_X\lambda^{\alpha}q(\Gamma^X)_{\alpha\beta}\lambda^{\alpha}_{ba},$$
with $\lambda^{\alpha}$ standing for $SU(3)$ color Gell-Mann matrices.
There are 16 basic $4\times 4$ matrices $\Gamma_{X,Y}$ acting on the Lorentz indices of the quark operators. They are
$\Gamma_X=I; \quad \Gamma_X=\gamma_5; \quad \Gamma_X=\gamma_{\mu}; \quad \Gamma_X=\gamma_{\mu}{\gamma_5}$,
with $\Gamma_X=i/2(\gamma_{\mu}\gamma_{\nu}-\gamma_{\nu}\gamma_{\mu})=\sigma_{\mu\nu} (\mu>\nu$.
They describe the scalar ($S$), pseudoscalar ($P$), vector ($V$), axial ($A$) and tensor ($T$) cases correspondingly.
Thus the four quark condensates have the Lorentz structure
$\langle M|\bar q\Gamma_Xq\bar q\Gamma_Yq|M\rangle.$
We do not display the color indices here.
We can write this expectation value as
$$ \langle 0| \bar q\Gamma_Xq\bar q\Gamma_Yq|0\rangle+\langle 0| \bar q\Gamma_Xq|0\rangle
\langle \tilde M|\bar q\Gamma_Yq|\tilde M\rangle+\langle \tilde M|\bar q\Gamma_Xq\bar q\Gamma_Yq|\tilde M\rangle.$$
The first term on the right hand side is the vacuum expectation value included in the vacuum sum rules.
The second (factorized) term describes the configuration in which one pair of quarks acts on the QCD vacuum while the other one
interacts with the nucleon matter. It will be referred to as the "nucleon term". We include the nucleon terms in which all four quarks act on the same nucleon of the matter. The configuration in which two pairs of quarks act on two different nucleons of the matter corresponds to of the $3N$ forces in QCD sum rules. We do not include them in the present paper.

The factorized terms obtain nonzero values for $SS$ and $SV$ structures, i.e. for $\Gamma_X=I$ and $\Gamma_Y=I$ or $\Gamma_Y=\gamma_0$.
Note that the scalar expectation value  $\langle N|\bar qq|N\rangle$ can be expressed in terms of observable nucleon sigma term.  The vector condensate is just $\langle N|\bar q\gamma_0q|N\rangle=n_q$. Thus the factorized terms can be obtained in a model independent way. Calculation of the nucleon terms shown in Fig ~1$e$ requires model assumptions on the nucleon structure.

As we said in Introduction, we employ a relativistic quark model suggested in \cite{12a}. The nucleon is considered as a system of three relativistic valence quarks moving in an effective static field. The valence quarks are supplemented by the pion cloud introduced with requirements of the chiral symmetry. The meson cloud is included in the lowest order of the perturbation theory. We employ the model in the version suggested in \cite{16} were the authors did not solve the Dirac equation for the given form of the effective field, but postulated the Gaussian shape of the constituent quark density.

Following the analysis carried out in \cite{X} we find that in the $\hat q$ structure the large factorized $SS$ term in almost totally canceled by contributions of the nucleon terms. The leading contribution to the $I$ structure is provided by the factorized $SV$ condensate. The $\hat P$ structure is dominated by the $VV$ and $AA$ nucleon terms.

Inclusion of the factorized $SV$ condensate corresponds to exchange by vector meson between the probe nucleon and the nucleons of the matter.
This exchange has an anomalous vertex since it contributes to the scalar self energy of the probe nucleon.
Note that such contributions can be cause by the Firtz transformed exchange terms of the vector interaction between the probe nucleon with the nucleons of the matter.
The nucleon term with $VV$ and $AA$ condensate corresponds to exchange by two mesons with corresponding quantum numbers.

The contributions of the four-quark condensates can be presented as
\begin{equation}
 A_6^{(m)}=8a\pi^2A_q\rho; \quad B_6^{(m)}=4\pi^2\frac{s-m^2}{m}aB_q\rho; \quad C_6^{(m)}=4\pi^2 \frac{s-m^2}{m}aC_q\rho,
\label{20}
\end{equation}
with $s=4m^2$, $a=-0.55 $GeV$^3$. The dimensionless parameters $A_q=-0.10$, $B_q=-1.80$ and $C_q=-0.88$ were obtained in \cite{X}.
Thus the contributions of the four-quark condensates diminish the values of $\Sigma_V$ and $m^*$ at $\rho=\rho_0$.
Hence the equilibrium states, if there are any manifest themselves at somewhat larger values of $\rho$.

Now the matter has equilibrium states for $\sigma_N \geq 41.2$~MeV -- see Fig.~7.
As in the previous case and for the same reasons there are no equilibrium states if we put  $F(\rho)=1$.

We find ${\cal B}_{min}=-16.04$~MeV for $\sigma_N =43.6$~MeV. The latter number is close to the conventual value $\sigma_N =45$~MeV. The corresponding equilibrium density is $\rho_{eq}=1.46\rho_0$.

In Fig.~8$a$ we present the results for the vector self energy
$\Sigma_V$, for the nucleon effective mass $m^*$. In Fig.~8$b$ we demonstrate the density dependence of the residue  $\lambda_m^2$ and that of the effective threshold $W_m^2$. The function $F$ is determined by Eq.~(\ref{7}). More detailed results for the values of density close to $\rho_{eq}$ are presented in Table 2. At the equilibrium point $\Sigma_S=-482$~MeV. Thus the values of $\Sigma_S$ at the equilibrium points of our approach and of the  Walecka model differ by about $70$~MeV. The value of the vector self energy $\Sigma_V=387$~MeV exceeds that of the Walecka model by $65$~MeV. The single-particle potential energy equilibrium value $U=-95$~MeV is very close
to the Walecka model result $U=-90$~MeV.

Equations (\ref{12}) enable us to write the physically motivated parametrization for the density dependence of nucleon parameters.Since the shift of the threshold $W_m^2-W_0^2$  is small, it can be treated perturbatively and can be assumed to be proportional to $\rho$ (see Fig.~8$b$). Start with expression for $\Sigma_V$.
Since ${\cal L}^P$ and ${\cal L}^q$ are linear in $\rho$ we write immediately
\begin{equation}
 \Sigma_V=\frac{a_Vx}{1+a_qx}; \quad x=\frac{\rho}{\rho_0},
\label{21}
\end{equation}
with $a_q,a_V$ the coefficients which chosen for the best fitting of the right hand side. Turning to expression for $m^{*}$ we see that ${\cal L}^I$
also contains the terms which are proportional to $\rho$. They are caused by the four-quark condensates and by the finite value of the shift $W_m^2-W_0^2$. However there are also the nonlinear terms which are proportional to the product $\rho F(\rho)$ caused by the scalar condensate. Thus we can
write
$$m^{*}=\frac{m+a_S^{(1)}x+a_S^{(2)}x^2}{1+a_qx}.$$
The second term in numerator approximates the nonlinear contribution.
We find that $a_q=-0.190$ and $a_V=186$~MeV for $\sigma_N =43.6$~MeV. The effective mass is well approximated by the choice $a_S^{(1)}=-468$~MeV;  $a_S^{(2)}=43$~MeV. The scalar self energy can be thus presented as
\begin{equation}
 \Sigma_S=\frac{\tilde a_S^{(1)}x+a_S^{(2)}x^2}{1+a_qx},
\label{22a}
\end{equation}
with $\tilde a_S^{(1)}=-292$~MeV.
One can see that $a_S^{(1)}$ is proportional to $\sigma_N$. The parameters $a_q$ and $a_V$ exhibit weak dependence on the value of  sigma term.

\begin{table}
\caption{Nucleon parameters obtained with inclusion of the condensates with $d \leq 6$ and of the lowest order radiative corrections,
$\sigma_N=43.6$~MeV.
For each value of the baryon density $\rho$ the upper line corresponds to $\kappa_N^{eff}$ determined by Eq.(\ref{9}).
The lower line is for $\kappa_N^{eff}=\kappa_N$.}

\begin{center}
\begin{tabular}{|c|c|c|c|c|c|}
\hline
$\rho/\rho_0$&$m^*$,~MeV&$\Sigma_V$, MeV&$\lambda_m^2$, GeV$^6$&$W_m^2$, GeV$^2$\\
\hline
1.40&484&351&1.23&1.70\\
&424&347&1.16&1.68\\
\hline
1.45&465&369&1.20&1.69\\
&396&366&1.13&1.67\\
\hline
1.50&446&387&1.18&1.68\\
&367&384&1.11&1.67\\
\hline
1.55&428&407&1.16&1.68\\
&336&404&1.09&1.66\\
\hline
1.60&411&427&1.15&1.67\\
&304&423&1.06&1.65\\
\hline
\end{tabular}
\end{center}
\end{table}
In Fig.~9 we present dependence of ${\cal B}_{min}$ on the value of $\sigma_N$. The dependence of $\rho_{eq}$ on $\sigma_N$ is shown in Fig.~10.

\section{Summary}
We calculated the nucleon parameters in symmetric nuclear matter focusing on the average energy per nucleon ${\cal B}$. We include only the contributions corresponding to the $2N$ forces. Analysis of the behavior of the energy ${\cal B}(\rho)$ enabled us to investigate the equilibrium states of the matter.

We use the finite density QCD sum rules method. In this approach the nucleon parameters are expressed in terms of the QCD condensates. The lowest dimension condensates ($d=3$) are the vector and scalar expectation values. Calculation of the scalar condensate requires certain assumptions on the structure of the matter. We treat the matter as a relativistic system of interacting nucleons. This differs from the previous QCD sum rules studies which start with considering of the polarization operator in the system of nonrelativistic free nucleons.

The Hellmann-Feynman theorem ties the scalar condensate with the derivative of the relativistic energy density with respect to the light quark mass.
We assume that the dependence of the energy density on the light quark mass is dominated by that of the nucleon mass. The contribution of the nucleon interaction is relatively small. The assumption is consistent with the result that the model calculations provide small contribution of nucleon interactions to the scalar condensate. This enables us to express the scalar condensate as the product of the nucleon matrix element $\kappa_N$ and
the function $F(\rho)$ which can be expressed in terms of the Lorentz factors of the nucleons of the matter -- Eq.~(18). The same function $F(\rho)$
ties the baryon and scalar densities in the scalar-vector model of nuclear matter \cite{10}. The nucleon matrix element $\kappa_N$ can be expressed
in terms of the observable nucleon sigma term $\sigma_N$.

We start with condensates of dimension $d=3$ adding consistently those of the higher dimension. We find that ${\cal B}(\rho)>0$ if only the condensates with $d=3$ are included. Thus there are no equilibrium states. Including the condensates with dimension $d=4$ and the lowest order radiative corrections we find the equilibrium states for the values of $\sigma_N$ near the upper limit of the interval determined by Eq.~(20). For $\sigma_N=65$~MeV the equilibrium density surprisingly close to the phenomenological value $\rho_0$, i.e. $\rho_{eq}=0.99\rho_0$. The average energy
per nucleon ${\cal B}(\rho_{eq})=-12.7$~MeV. The values of $m^*$ and $\Sigma_V$ are close to those provided by the Walecka model at its saturation point.

The contributions of the condensates with $d \leq 4$ are obtained in a model independent way. The calculation of the four-quark condensates ($d=6$)
requires model assumptions on the quark structure of the nucleon. We employ the relativistic quark model suggested in \cite{12a},\cite{16}. Since we
consider only the $2N$ contributions, we include only the configurations in which all four quark operators act on the same nucleon of the matter.
We find equilibrium states of the matter for all $\sigma_N >41$~MeV. The equilibrium with ${\cal B}(\rho_{eq})=-16.04$~MeV is realized at $\sigma_N =43.6$~MeV, very close to the conventional value. For this value of the sigma term $\rho_{eq}=1.46\rho_0$. The values of $m^*$ and $\Sigma_V$ differ by
about 20 percent from those given by the Walecka model. However the values of the single-particle potential energies $U(\rho_{eq})$ in the two approaches are very close.

Note that putting $F(\rho)=1$ we find no equilibrium states. Thus the saturation is an essentially relativistic effect in our approach, similar to
the Walecka model.

To find the contribution of the $3N$ interactions we must include the configurations of the four-quark condensates where two pairs of quark operators act on two different nucleons of the matter. This will be a subject of next work.

{}

\newpage

\begin{figure}
\centering{\epsfig{figure=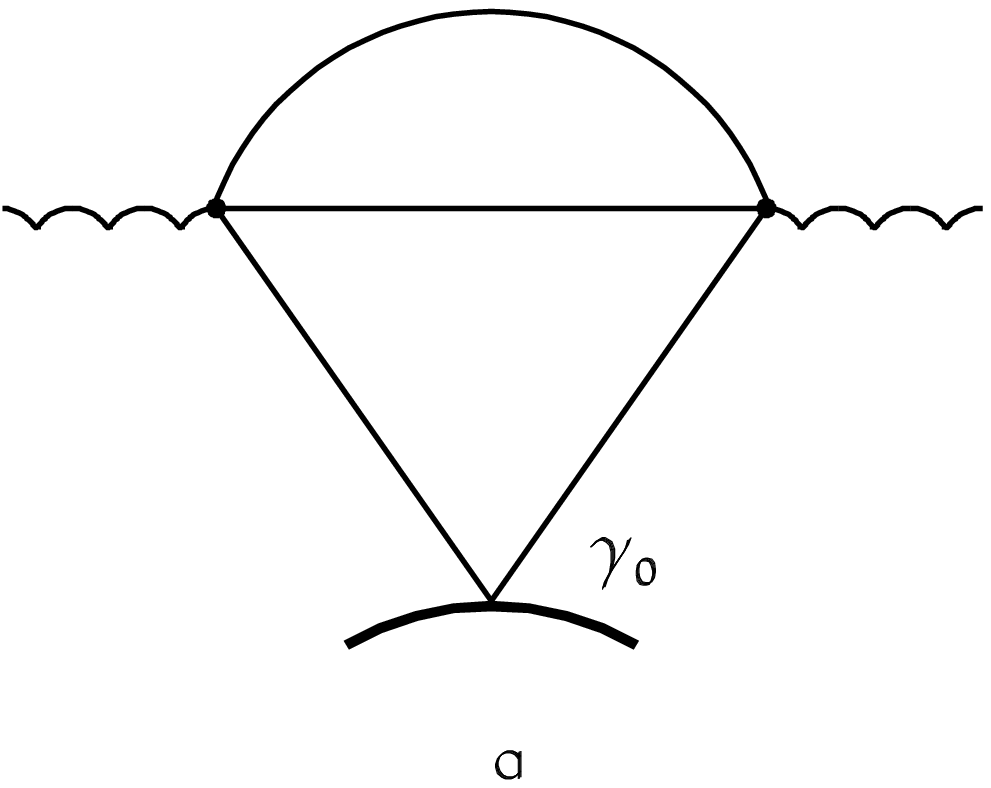,width=6cm}\hspace{0.5cm}\epsfig{figure=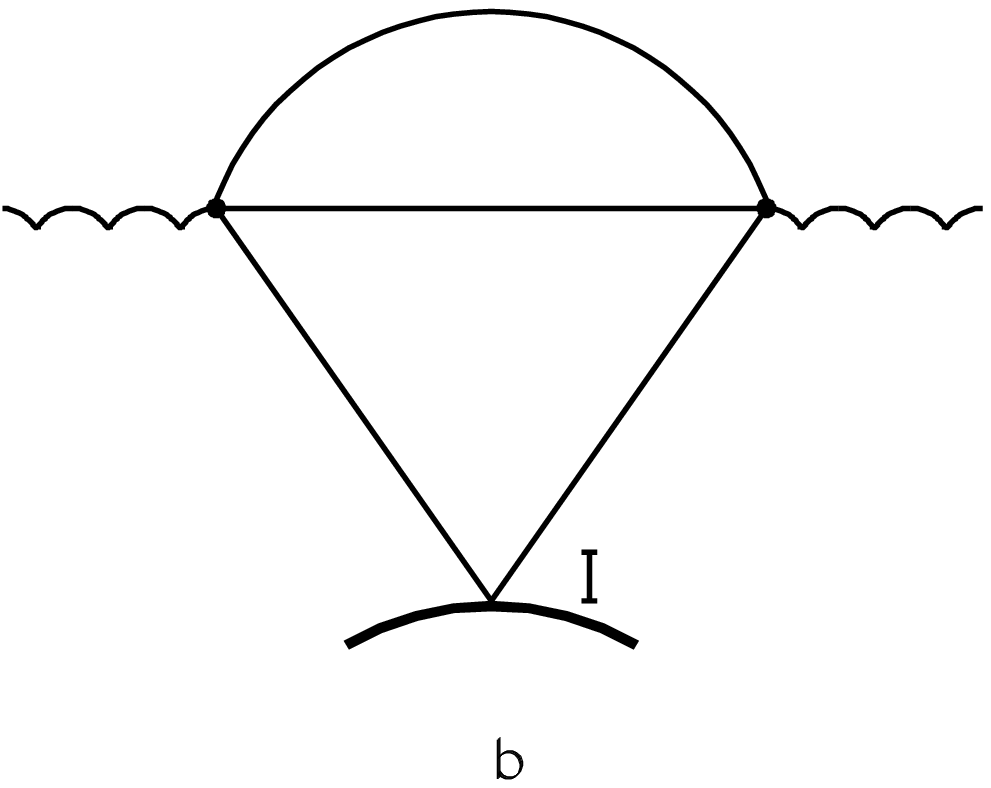,width=6cm}}

\vspace{-0.5cm}

\centering{\epsfig{figure=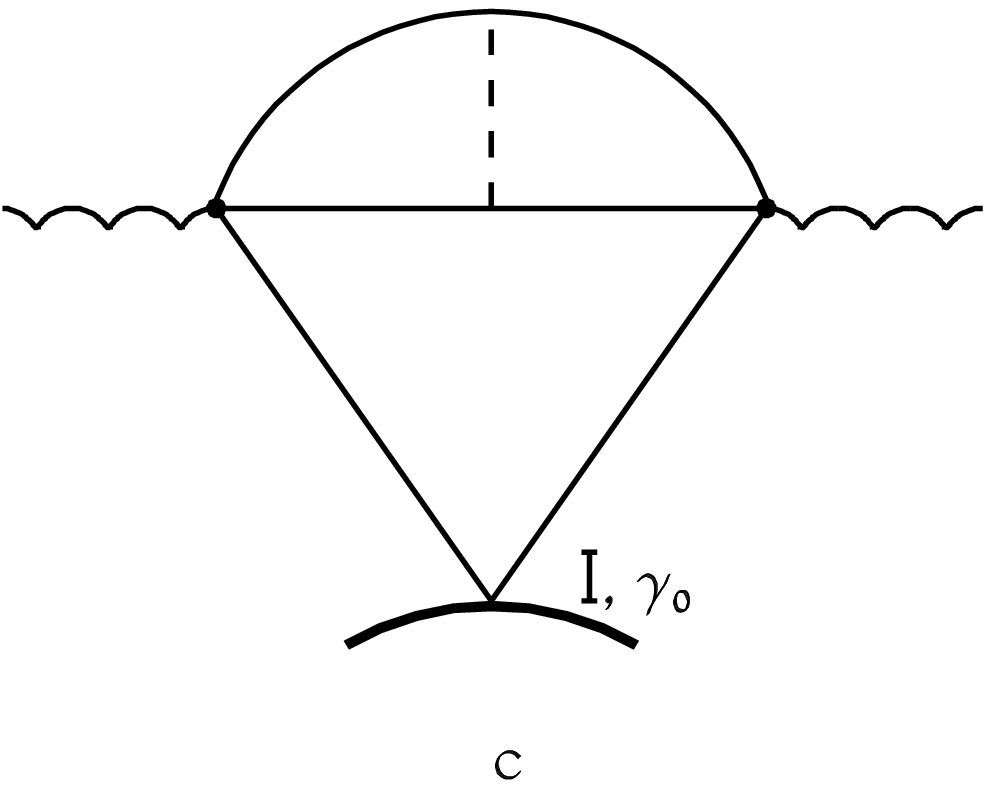,width=6cm}\hspace{0.5cm} \epsfig{figure=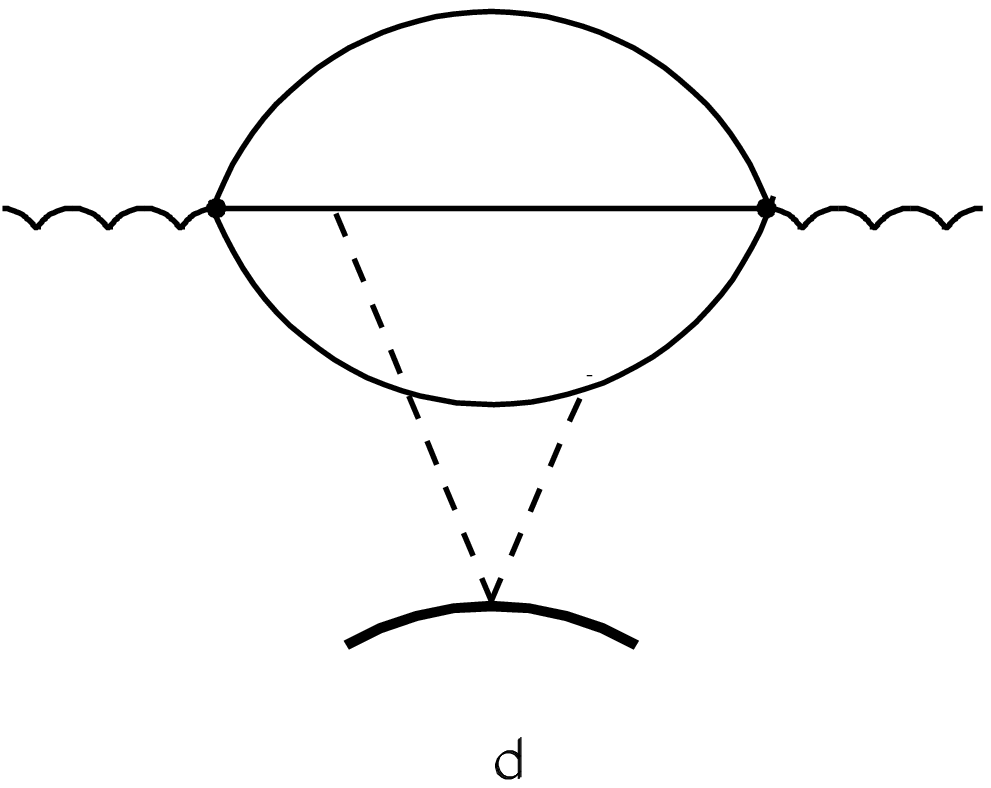,width=6cm}}

\vspace{-0.5cm}

\centering{\epsfig{figure=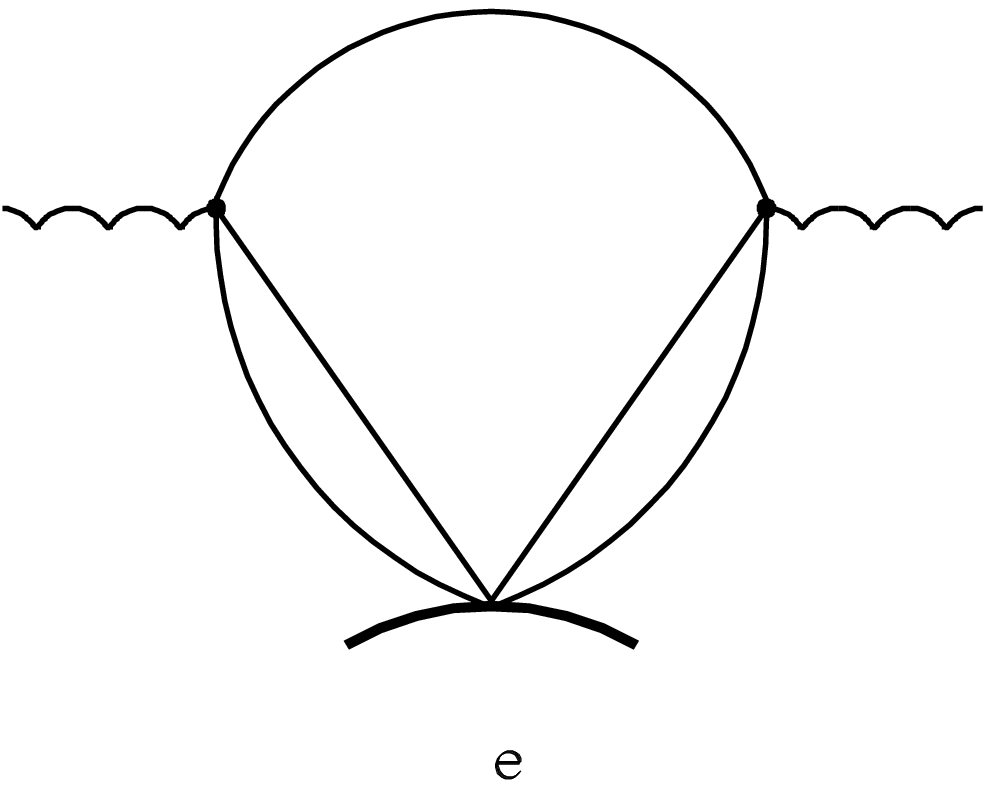,width=6cm}}
\caption{Feynman diagrams for the leading OPE terms of polarization operator. The helix line denotes the system with the quantum numbers of the proton. Solid lines stand for the quarks. The bold lines are for the nucleons of the matter. Dashed lines denote gluons. Figs.~$a,b$ -- contributions of the vector and scalar condensates; Figs.~$c - a$ typical in-medium radiative correction; Fig.~$d$ -- contribution of the gluon condensate; Fig.~$e$ -- contributions of the four-quark condensates.}
\label{f1}
\end{figure}

\begin{figure}
\centering{\epsfig{figure=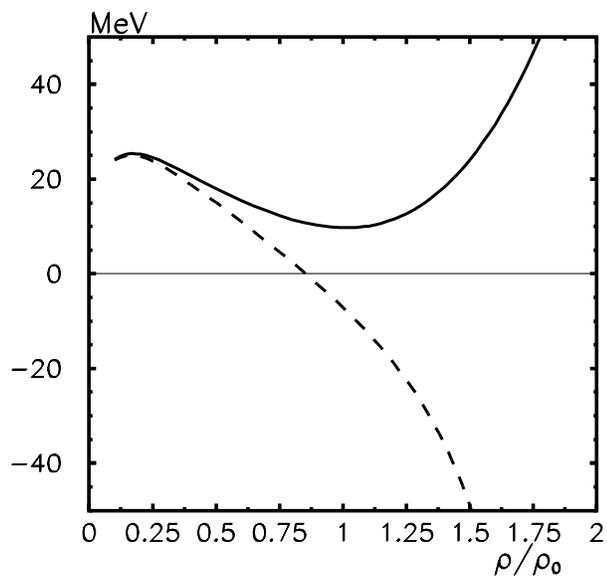,width=9cm}}
\caption{ Density dependence of the average binding energy per nucleon ${\cal B}$ with only the condensates with dimension $d=3$ included; $\sigma_N=65$~MeV. The vertical axis is for ${\cal B}(\rho)$ in MeV.
For each value of the baryon density $\rho$ the upper line corresponds to $\kappa_N^{eff}$ determined by Eq.(\ref{9}).
The lower line is for $\kappa_N^{eff}=\kappa_N$.}
\label{f2}
\end{figure}

\begin{figure}
\centering{\epsfig{figure=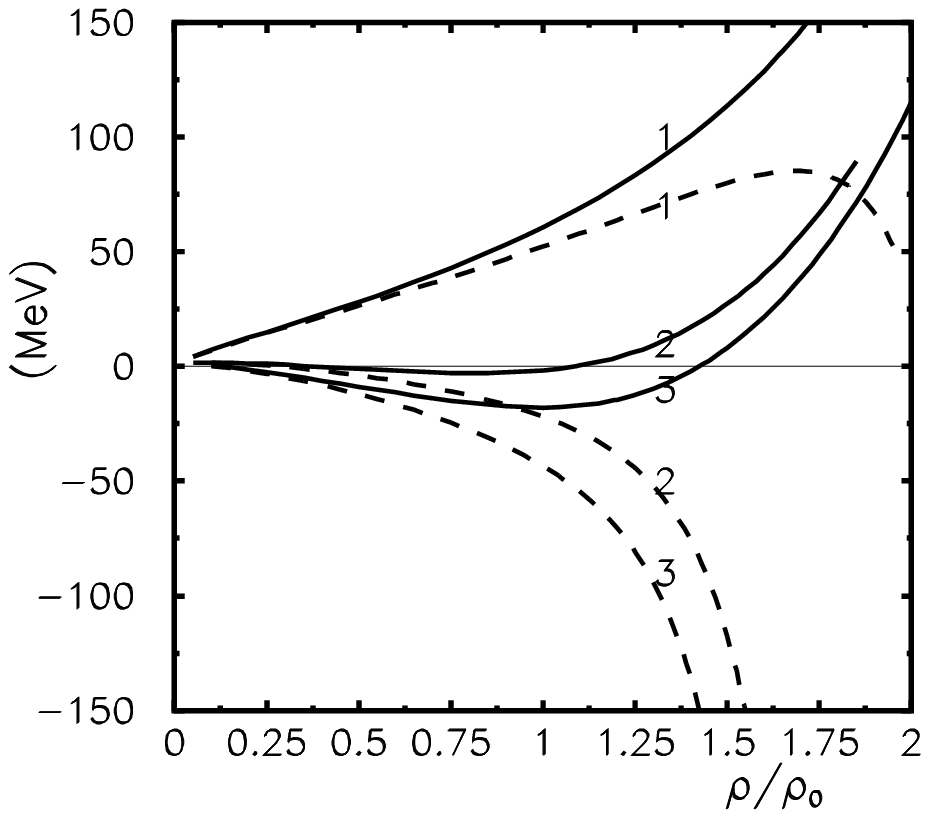,width=9cm}}
\caption{Density dependence of the average binding energy per nucleon ${\cal B}$. The vertical axis is for ${\cal B}(\rho)$ in MeV.
The radiative corrections and the condensates with dimensions $d\leq 4$ are included. The numbers 1,2,3 are for the cases $\sigma_N$=45, 61, 65~MeV,  correspondingly. Other notations are the same as in Fig.~2.}
\label{f3}
\end{figure}

\begin{figure} 
\centering{\epsfig{figure=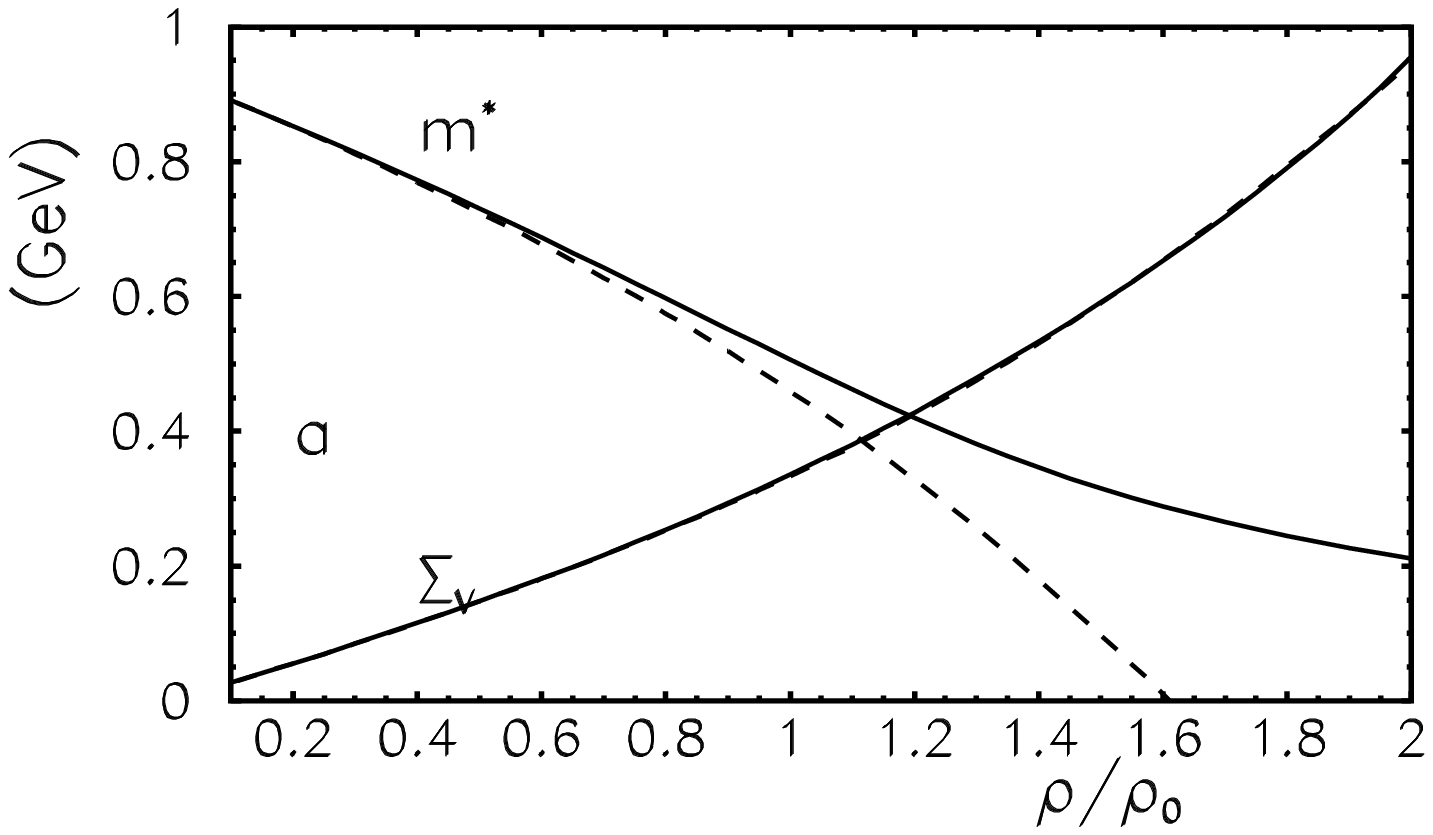,width=9cm} \epsfig{figure=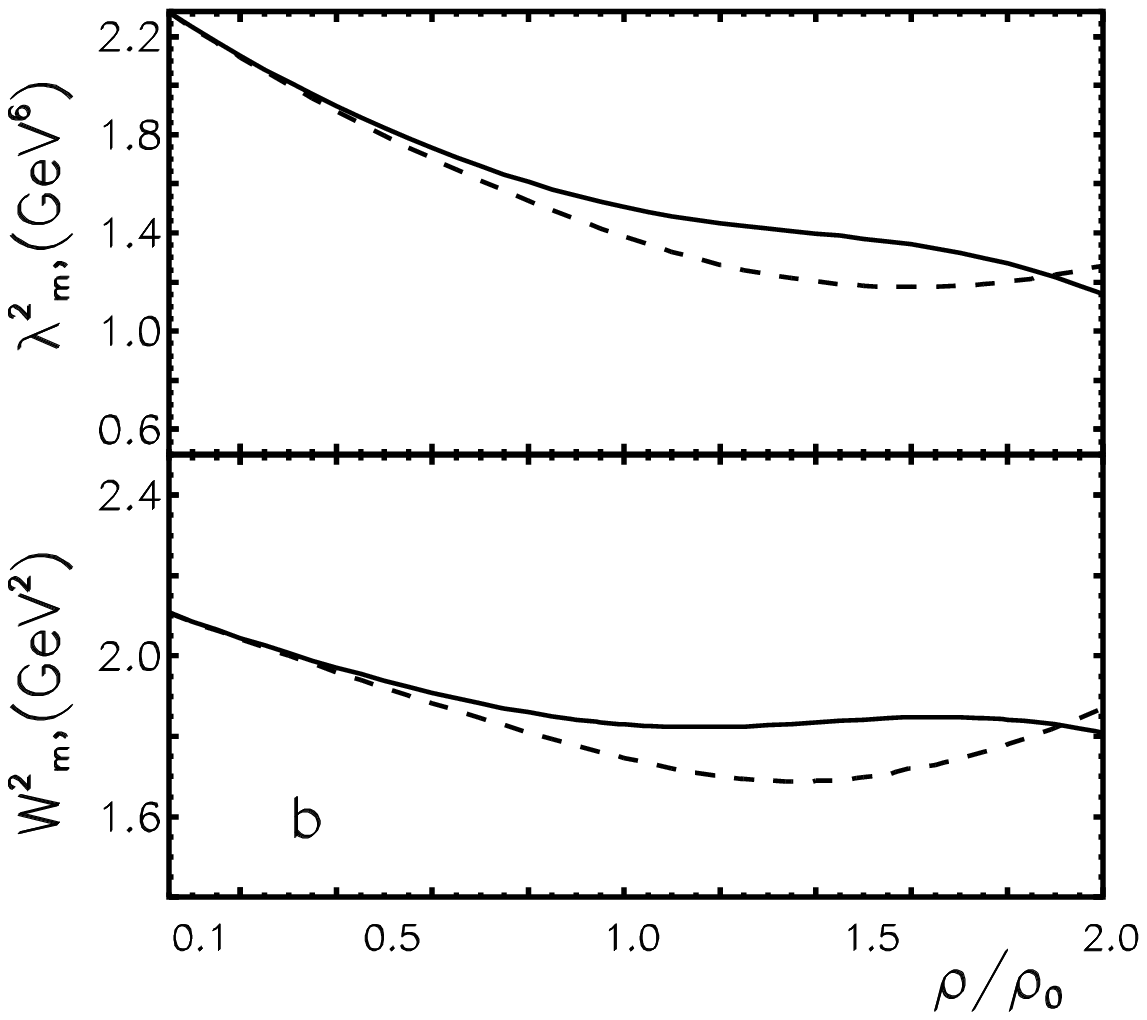,width=9cm}}
\caption{ Density dependence of the nucleon parameters. The radiative corrections and the condensates with dimensions $d\leq 4$ are included. The horizontal axis corresponds to the density $\rho$ related to its empirical saturation value $\rho_0$. Fig.~$a$ -- effective nucleon mass $m^*$ and of the vector self energy $\Sigma_V$. Fig.~$b$ -- the nucleon residue $\lambda_m^2$ and  the continuum threshold $W_m^2$. The meaning of the solid and  dashes lines is the same as  in Fig.~2.}
 \label{f4}
\end{figure}

\begin{figure}
\centering{\epsfig{figure=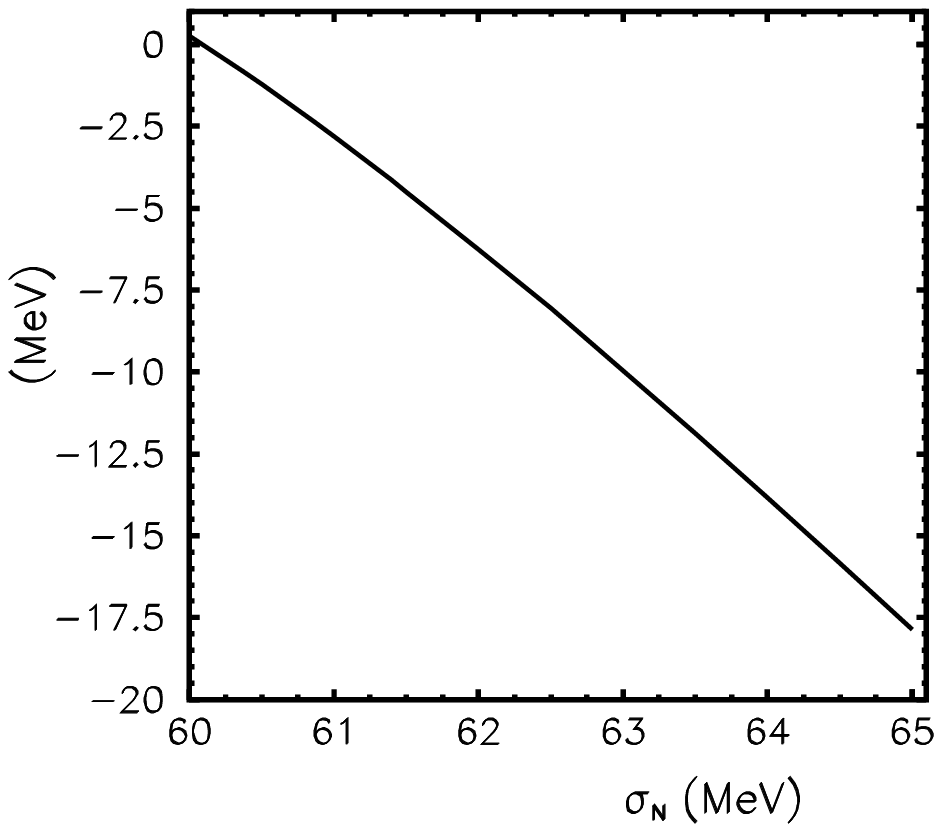,width=8cm}}
\caption{ Dependence of the average binding energy per nucleon ${\cal B}(\rho_{eq})$ in equilibrium state of the matter on the value of $\sigma_N$. The vertical axis is for ${\cal B}(\rho_{eq})$ in MeV. The radiative corrections and the condensates with dimensions $d\leq 4$ are included.}
\label{f5}
\end{figure}

\begin{figure}
\centering{\epsfig{figure=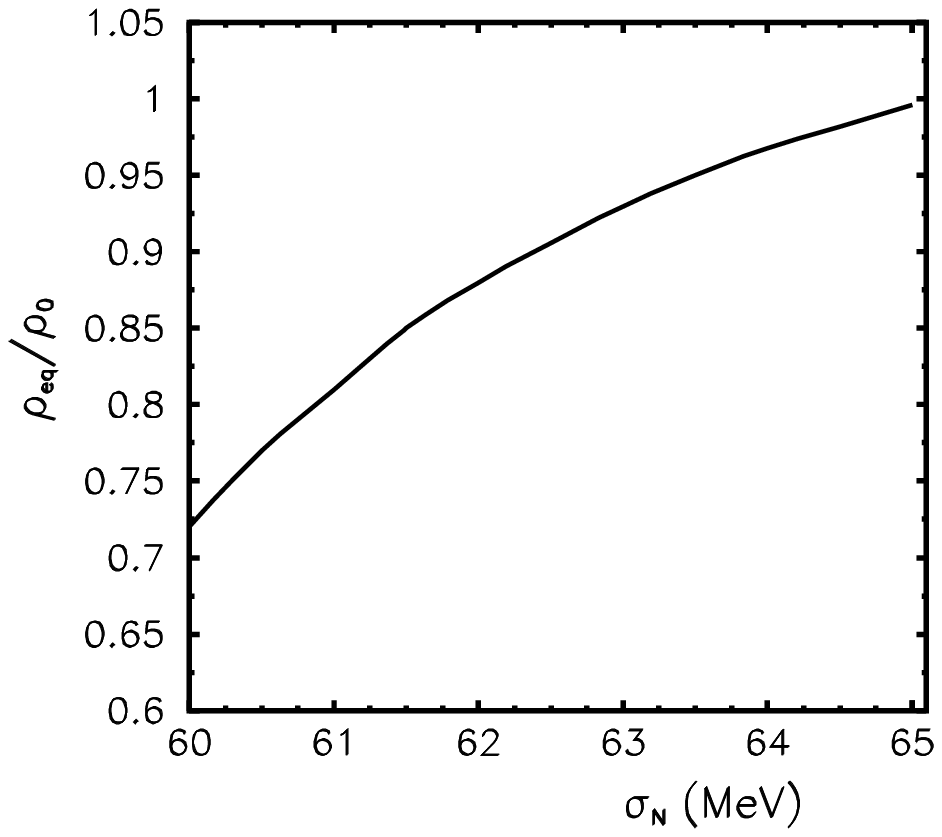,width=8cm}}
\caption{Dependence of the equilibrium density $\rho_{eq}$ on the value of $\sigma_N$. The radiative corrections and the condensates with dimensions $d\leq 4$ are included.}
\label{f6}
\end{figure}

\begin{figure}
\centering{\epsfig{figure=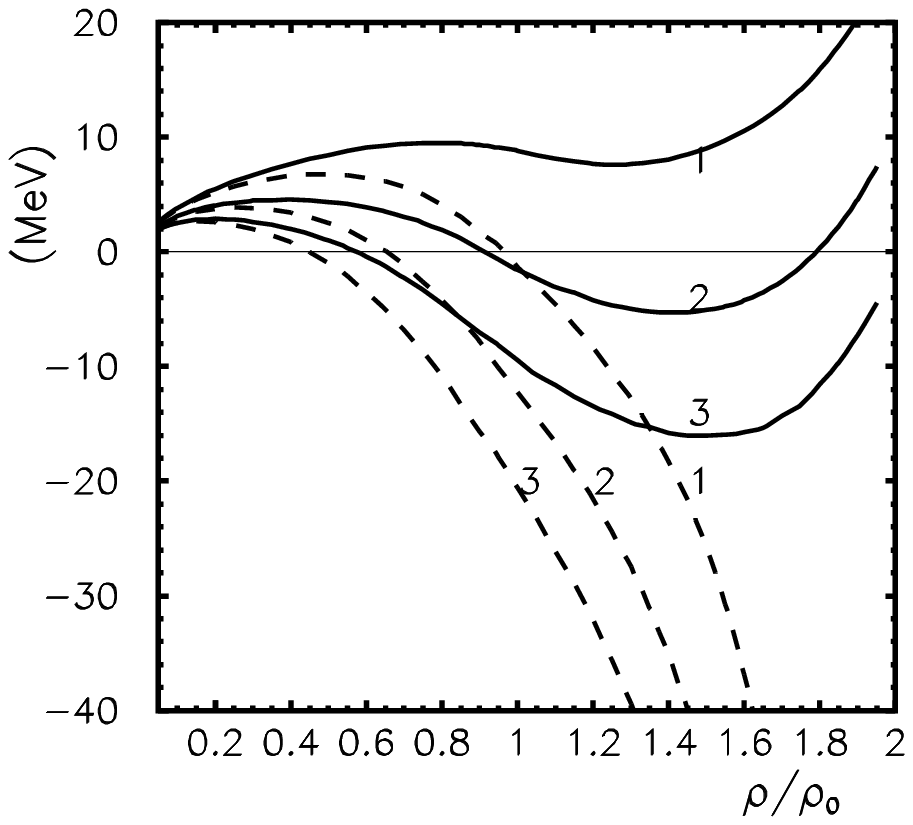,width=9cm}}
\caption{Density dependence of the average binding energy per nucleon ${\cal B}$. The vertical axis is for ${\cal B}(\rho)$ in MeV.
The radiative corrections and the condensates with dimensions $d\leq 6$ are included. The numbers 1,2,3 are for the cases $\sigma_N$=40, 42, 43.6~MeV, correspondingly. Other notations are the same as in Fig.~2.}
\label{f7}
\end{figure}

\begin{figure} 
\centering{\epsfig{figure=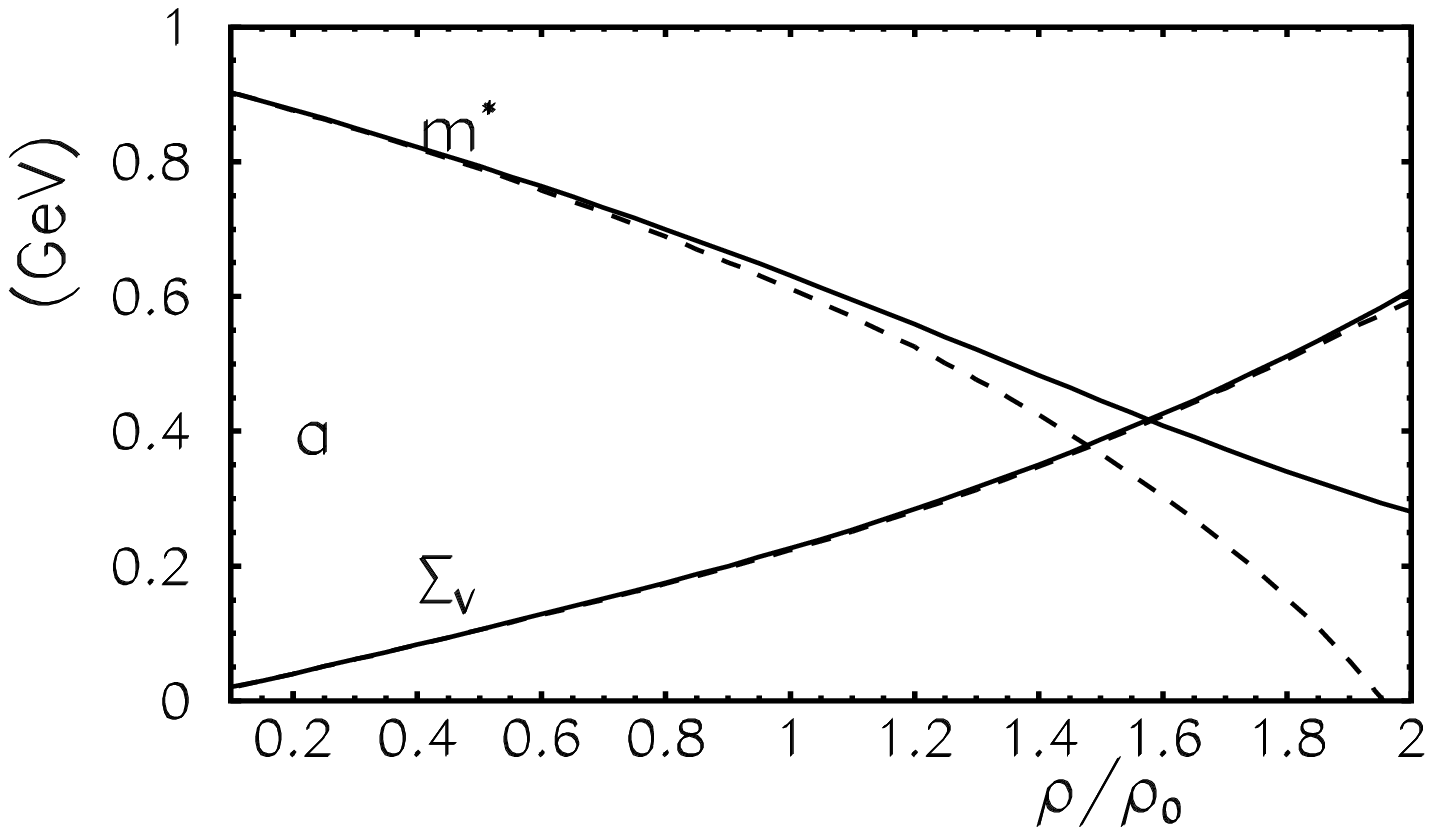,width=9cm} \epsfig{figure=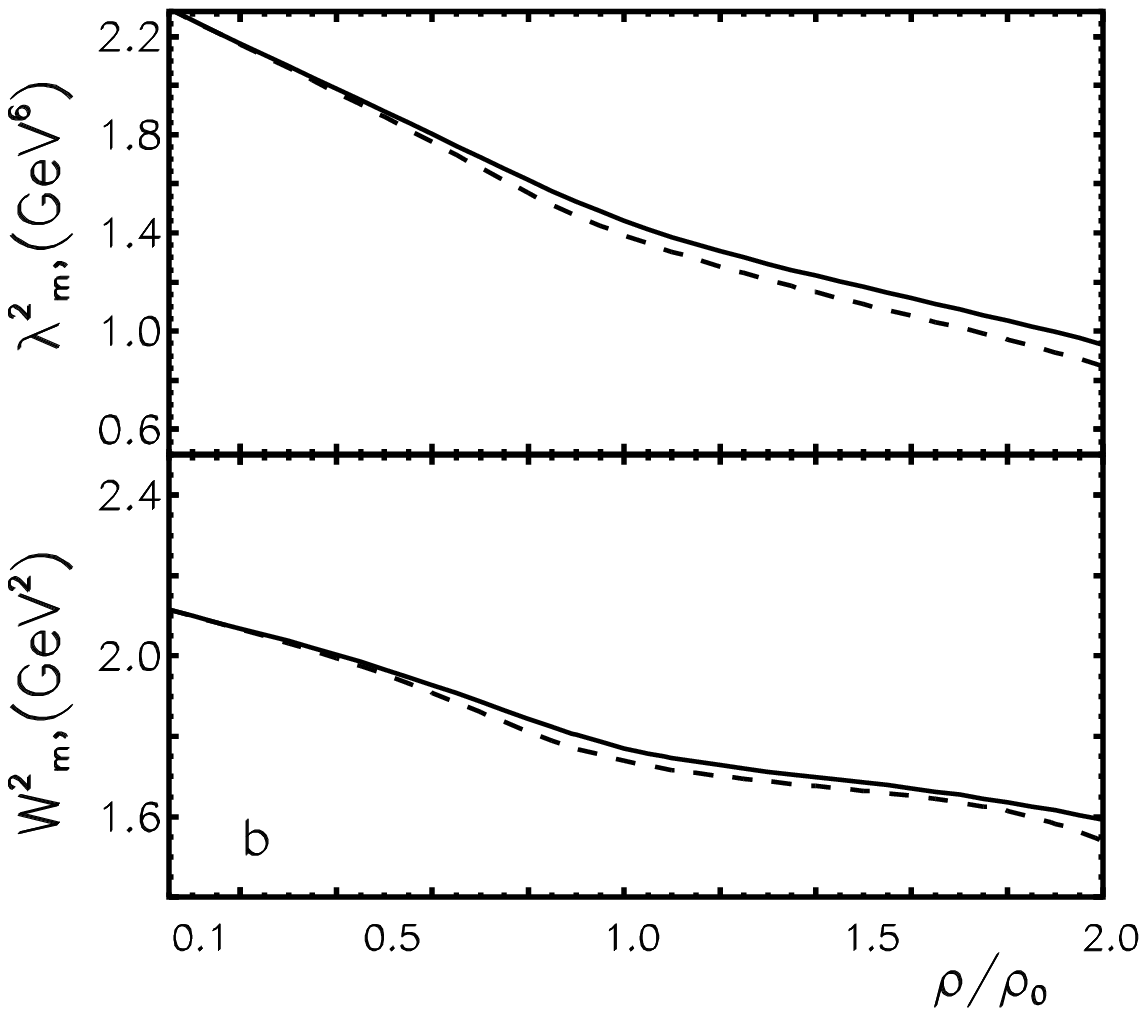,width=9cm}}
\caption{Density dependence of the nucleon parameters. The radiative corrections and the condensates with dimensions $d\leq 6$ are included. The horizontal axis corresponds to the density $\rho$ related to its empirical saturation value $\rho_0$. Fig.~$a$ -- effective nucleon mass $m^*$ and of the vector self energy $\Sigma_V$. Fig.~$b$ -- the nucleon residue $\lambda_m^2$ and of the continuum threshold $W_m^2$. The meaning of the solid and  dashes lines is the same as  in Fig.~2.}
 \label{f8}
\end{figure}

\begin{figure}
\centering{\epsfig{figure=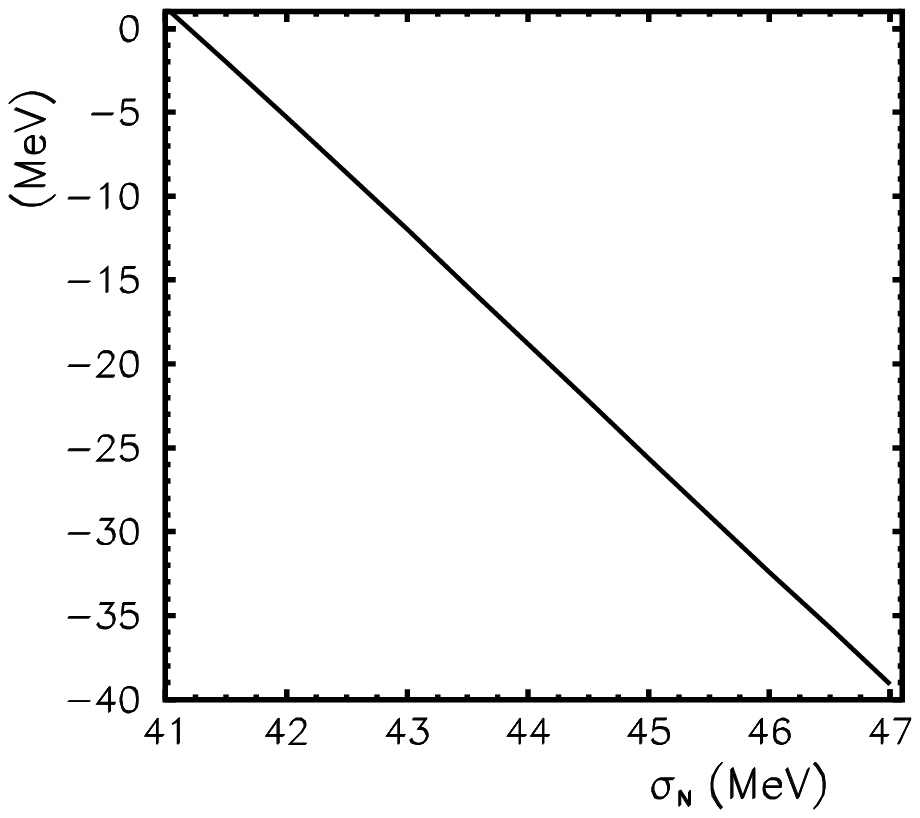,width=8cm}}
\caption{Dependence of the average binding energy per nucleon ${\cal B}(\rho_{eq})$ in equilibrium state of the matter on the value of $\sigma_N$.
The vertical axis is for ${\cal B}(\rho_{eq})$ in MeV. The radiative corrections and the condensates with dimensions $d\leq 6$ are included.}
\label{f9}
\end{figure}

\begin{figure}
\centering{\epsfig{figure=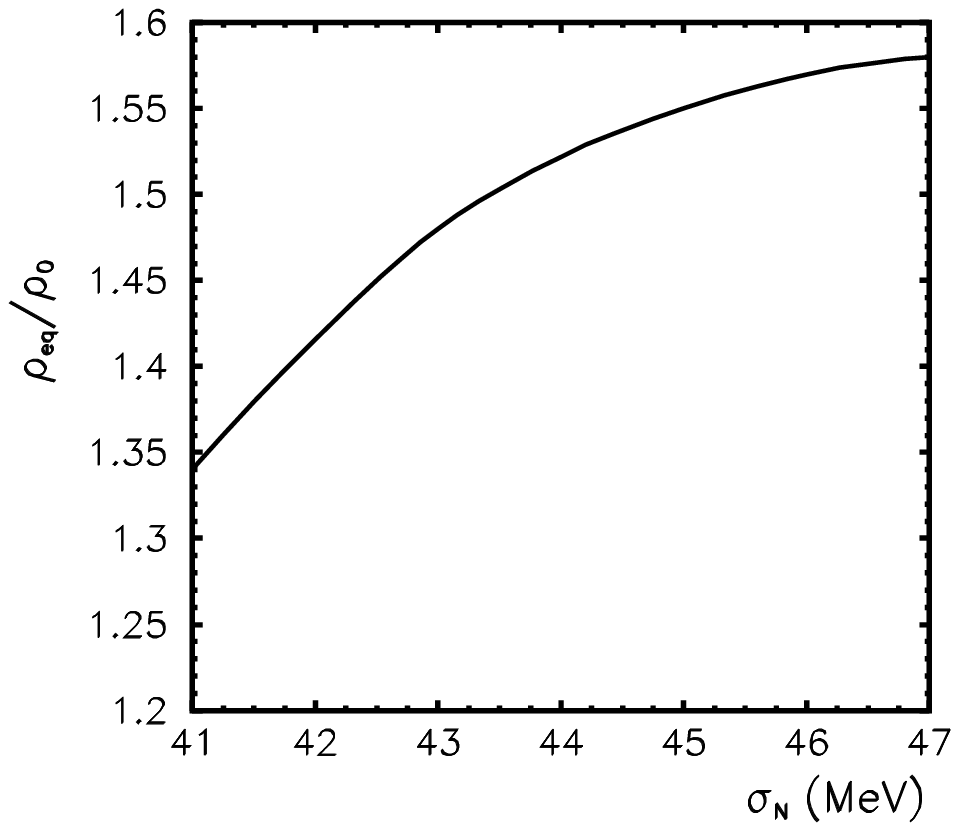,width=8cm}}
\caption{ Dependence of the equilibrium density $\rho_{eq}$ on the value of $\sigma_N$. The radiative corrections and the condensates with dimensions $d\leq 6$ are included.}
\label{f10}
\end{figure}

\end{document}